\documentclass[conference]{IEEEtran}
\IEEEoverridecommandlockouts
\usepackage{cite}
\usepackage{amsmath,amssymb,amsfonts}
\usepackage{graphicx}
\usepackage{textcomp}
\usepackage{xcolor}
\usepackage{subcaption}
\usepackage{float}
\usepackage{algorithm}
\usepackage{algorithmic}

\DeclareMathOperator*{\argmax}{arg\,max}

\def\BibTeX{{\rm B\kern-.05em{\sc i\kern-.025em b}\kern-.08em
    T\kern-.1667em\lower.7ex\hbox{E}\kern-.125emX}}

\makeatletter
\def\ps@IEEEtitlepagestyle{%
\def\@oddfoot{\mycopyrightnotice}%
\def\@evenfoot{}%
}
\def\mycopyrightnotice{%
{\footnotesize 979-8-3503-8542-7/24/\$31.00~\copyright~2024 IEEE\hfill} 
\gdef\mycopyrightnotice{}
}

\begin{document}

\title{Bellman Memory Units:  A neuromorphic framework for synaptic reinforcement learning with an evolving network topology\\
 \thanks{Research Ireland Centre for Research Training in Foundations of Data Science.}
 }

\author{\IEEEauthorblockN{Shreyan Banerjee\IEEEauthorrefmark{1},
Aasifa Rounak\IEEEauthorrefmark{2}, Vikram Pakrashi\IEEEauthorrefmark{3}}
\IEEEauthorblockA{School of Mechanical and Materials Engineering, UCD Centre for Mechanics, Dynamical Systems and Risk Laboratory \\
University College Dublin\\
Dublin, Ireland \\
Email: \IEEEauthorrefmark{1}shreyan.banerjee@ucdconnect.ie,
\IEEEauthorrefmark{2}aasifa.rounak@ucd.ie,
\IEEEauthorrefmark{3}vikram.pakrashi@ucd.ie}}

\maketitle

\begin{abstract}
Application of neuromorphic edge devices for control is limited by the constraints on gradient-free online learning and scalability of the hardware across control problems. This paper introduces a synaptic Q-learning algorithm for the control of the classical Cartpole, where the Bellman equations are incorporated at the synaptic level. This formulation enables the iterative evolution of the network topology, represented as a directed graph, throughout the training process. This is followed by a similar approach called neuromorphic Bellman Memory Units (BMU(s)), which are implemented with the Neural Engineering Framework on Intel's Loihi neuromorphic chip. Topology evolution, in conjunction with mixed-signal computation, leverages the optimization of the number of neurons and synapses that could be used to design spike-based reinforcement learning accelerators. The proposed architecture can potentially reduce resource utilization on board, aiding the manufacturing of compact application-specific neuromorphic ICs. Moreover, the on-chip learning introduced in this work and implemented on a neuromorphic chip can enable adaptation to unseen control scenarios.
\end{abstract}

\begin{IEEEkeywords}
control, Loihi, Nengo, neuromorphic, reinforcement learning,  synapse, topology
\end{IEEEkeywords}

\section{Introduction}
\label{sec:introduction}
Neuromorphic computing \cite{b1} has evolved over the last three decades and has shown its usefulness in edge devices for real-time applications \cite{b1,b4}. Spiking neural networks (SNNs) form the building blocks of a neuromorphic architecture. The fundamental difference between analog neural networks (ANNs) and  SNNs lies in the handling of spatio-temporal data as sparsely distributed spikes of electrical voltage or current, as opposed to continuous signals \cite{b1}. This approach of SNNs can lead to low power and low latency computations \cite{b2,b5}, which are important considerations for edge devices. While neuromorphic sensing and perception \cite{b3} are widely studied, work on neuromorphic control is still limited, one of the reasons being the challenge of scalability of a chip across different control problems, due to an inadequate understanding of how conventional and data-driven control can benefit from neuromorphic devices \cite{b5}.

This limitation can be attributed to:
\begin{itemize}
 \item \textcolor{black}{Lack of neural computation primitives from sensors to processors to controllers, that can completely substitute their digital counterparts \cite{b5}}.
 \item Challenge of scalability for a chip across different control problems, due to inadequate knowledge about \textcolor{black}{connection of intelligence with dynamical neural models.\cite{b5}}
 \item Challenge of full neuromorphic integration spanning sensors to motor controllers (as discussed by Bartolozzi et. al \cite{b5}).
\end{itemize}
 
The current state-of-the-art for embodied intelligence, both in sensing and robotic control, along with the possibilities of full-scale neuromorphic integration on hardware, by replacing static frame-based computation with dynamic event-based computation, is discussed in \cite{b4,b5}. These works also discuss the importance of temporal multiplexing, both of capturing the temporal dynamics of systems and simulating the spatial connectivity of biological neurons on silicon.

Unlike conventional artificial neural networks, which implement synapses as a summing junction for the incoming inputs to a neuron,
neuromorphic architectures deploying spiking neurons use the synapse \cite{b16} as a digital-to-analog converter. SNNs leverage the functionality of a synapse beyond just being a summing junction. Here, incoming spikes are converted to an analog current signal by convolving the spike train with a non-linear kernel function \cite{b16}. For training spiking neural networks, learning rules are implemented in the synapse. Safa \textit{et. al.} \cite{b17} surveys Hebbian and Spike Time Dependent Plasticity (STDP) unsupervised learning rules for sparse and predictive coding in neuromorphic architectures. 

In addition to Hebbian learning rules, reward-influenced learning, called reinforcement learning (RL), is another potential candidate for learning in neuromorphic systems.
Literature shows evidence of RL being used at the synaptic level. For example, the use of RL to modify synaptic weights for learning is shown in \cite{b6}.  
Qiu, \textit{et. al.}\cite{b7} uses the NEAT algorithm for network evolution with SNNs for balancing a cartpole. Lele, \textit{et al.}\cite{b8} uses rewards to modify the weights of a central pattern generator, to control a hexapod robot.
A comparative study between DQN and spiking DQN implemented on the Loihi neuromorphic chip to control a cartpole and an acrobot is shown in \cite{b10}.
Kaplanis \textit{et. al.} \cite{b18} shows a merger between the conventional Q-table approach and deep Q-networks using the Benna-Fusi biological neuron model with the main focus on reducing catastrophic forgetting in RL.
Considering such a diverse scope for new methods and hardware, an interesting proposition would be to combine a synapse-based RL implementation with mixed-signal architecture and evolving topology to solve a control problem, which may be deployed for control on the edge. 

This paper provides a preliminary example of such an approach and addresses the limitation of gradient-based learning by using reward-based learning at the synaptic level along with an evolving neural architecture to balance a pole on a moving cart. This is followed by an implementation of a similar algorithm (named as Bellman Memory Unit) on the Nengo Neural Engineering Framework \cite{b14}, and on Intel's Loihi neuromorphic chip \cite{b15}. 

The paper has been organized as follows. Section \ref{concepts} discusses Q-learning, and the cartpole simulator. 
Section \ref{algorithm} contains the proposed algorithm and its pseudo-code ,and section \ref{training} gives some training details. Results from such implementation have been outlined in Section \ref{results}, and finally, Section \ref{conclusion} discusses the conclusion and some future directions.

\section{Concepts}\label{concepts}

\subsection{Neuromorphic computing}
Neuromorphic computing is characterised by the use of spiking neurons and synapses to encode, store, and process information as temporally structured spike trains. This inherently parallel architecture offers significant advantages in power efficiency and scalability. Each synapse low-pass filters the incoming digital spike trains from pre-synaptic neurons, through multiple dendrites \cite{b9,b16}. It then produces a continuous analog time-signal at its output. The learning rules are implemented on the synapse. A neuromorphic architecture leverages the promise of providing low power, low latency and high scalability, making them highly suitable for edge devices to address perception and control application. 
 
A synapse, modeled as an RC series circuit \cite{b9}, convolves the input with a kernel, as seen in (\ref{syncurrent}).

\begin{equation}\label{syncurrent}
I_{syn}(t) = \Sigma_{x=-\infty}^{\infty} I_{in}(x) e^{-(\frac{t-x}{\tau})}
\end{equation}
where, $I_{syn}(t)$ is the post-synaptic current at time $t$ after the pre-synaptic spike train $I_{in}$ is filtered at the synapse, and $\tau$ is time constant of the synaptic kernel with which  $I_{in}$ is convolved. 
From Equation \ref{syncurrent}, a synapse convolves the input spike train with a kernel of time constant \(\tau_{syn}\) to produce an analog post-synaptic current. 
Such a synapse implementing a non-linear kernel function can be used to implement learning rules for training an SNN. One such rule is RL.
SNNs can be used to train RL models to be applied to control problems. 
A model-free, off-policy RL algorithm called Q-learning is chosen as it is easy to implement and requires only one neural network. This concept of Q-learning is briefly discussed in the next section.

\setlength{\textfloatsep}{0pt}

\subsection{Q-Learning}
\begin{figure}[htb]
\centering 
\includegraphics[clip=true, trim = {0cm 0 0cm 0cm}, width = 0.3\textwidth]{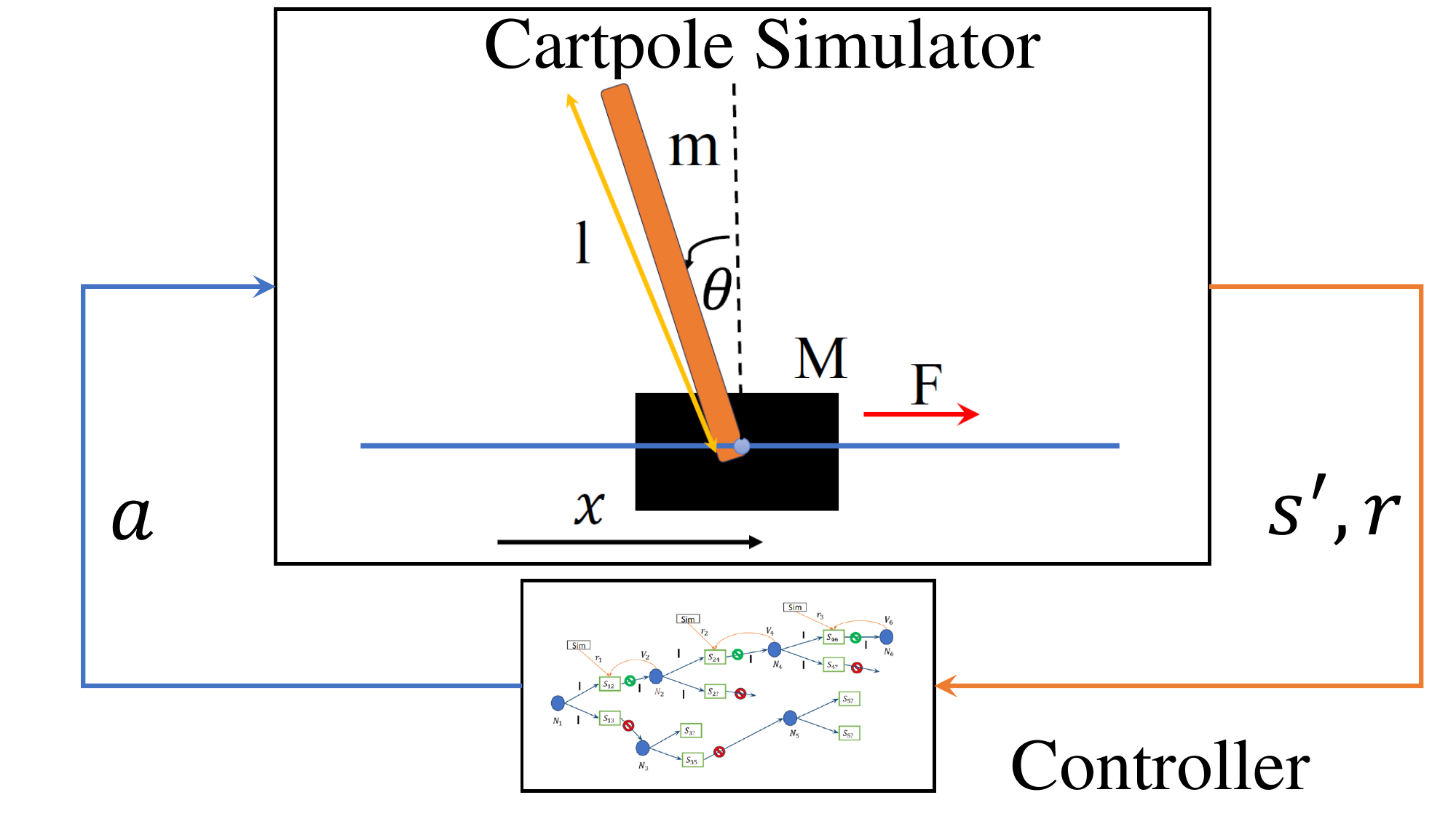}
\caption{A control system block diagram with synaptic RL-based controller in the loop. The simulator contains the cartpole model, where, $\text{M}$ is the mass of the cart, m is the mass of the pole and l is the length of the pole.} 
\vspace{3mm}
\label{fig:BD}
\end{figure}

Q-learning \cite{b10, b11} uses a scalar feedback signal called reward ($r$) fed back from the system, or environment, to the controller, or agent, to optimize its policy ($\pi(s)$) (see Fig. \ref{fig:BD}). The agent receives the state ($s \in \mathcal{S} $) from the environment and sends an action ($a \in \mathcal{A}(s)$) to the environment, such that the expected cumulative future reward ($Q(s,a)$) is maximized. This is known as an $\epsilon$-greedy approach since the agent selects actions greedily from $\pi(s)$, without allowing for any randomness or exploration.

The iterative update rule for $Q(s,a)$ is given by the Bellman equations (\ref{BellmanAlt}) and (\ref{value}).
\vspace*{-0.2cm} 
\begin{align}
Q(s,a) &:= Q(s,a) + \alpha (-Q(s,a) + r + \gamma V(s^{\prime})), \label{BellmanAlt}\\
V(s^{\prime}) &= \max_{a^{\prime}}Q(s^{\prime},a^{\prime}) \label{value}
\end{align}

Here, $(\cdot)^\prime$ represents variable $(\cdot)$ corresponding to the next state of the system, and $\alpha$ is the learning rate. $V(s^{\prime})$ is called the value function for the next state. The discounting factor $\gamma$ decides the future time until which the cumulative rewards are gathered. The lower the value of $\gamma$, the more short-sighted the agent.

In this work, RL-based control has been applied to the classic cartpole-balancing problem, as discussed in the subsequent sections. 

\begin{figure}[htb]
\centering 
\includegraphics[width = 0.45\textwidth]{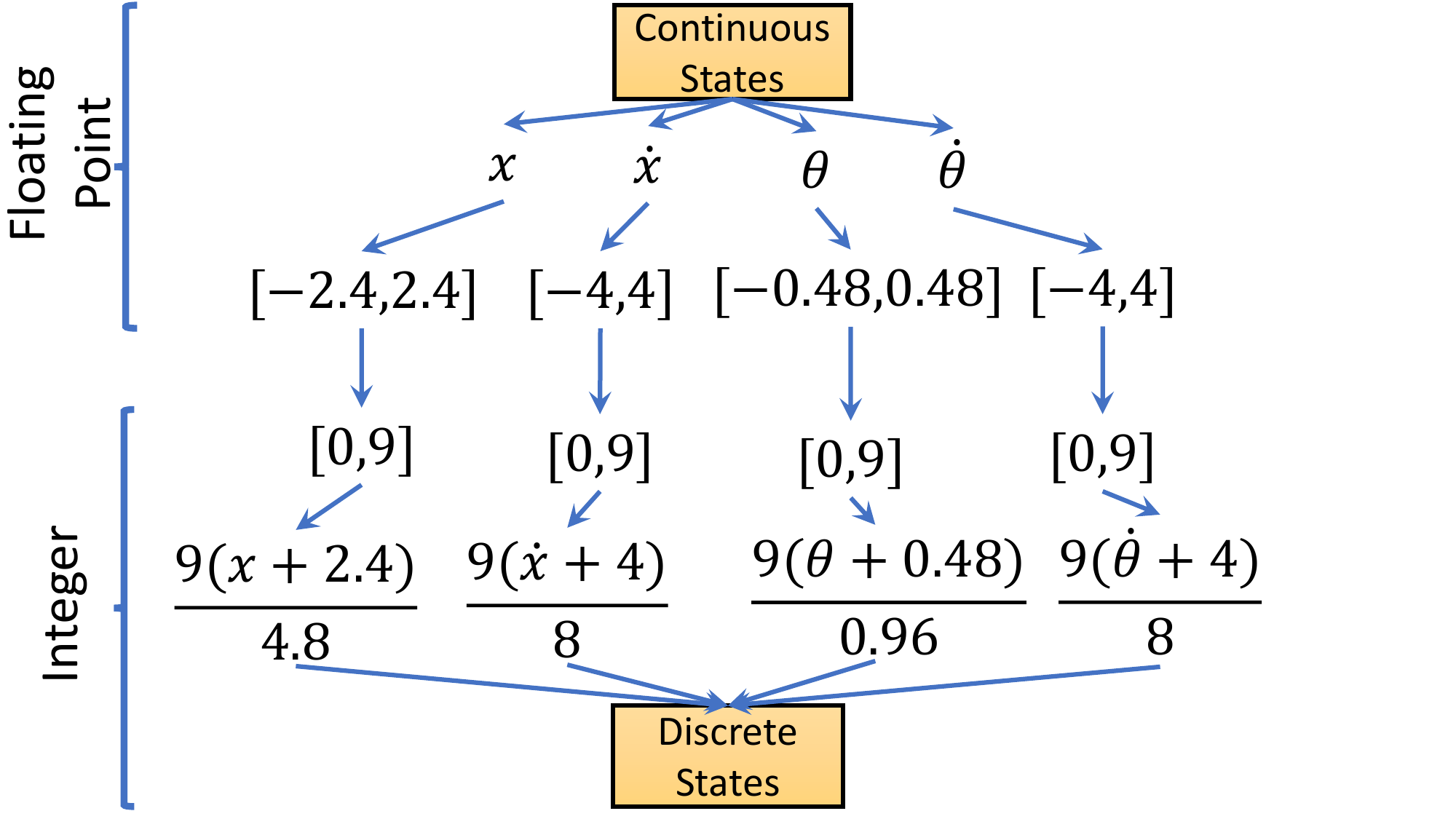}
\caption{Flow diagram showing the discretization of the cartpole state space.} 
\vspace{3mm}
\label{fig:DisFlow}
\end{figure}

\subsection{The Control Objective} \label{system}
Figure \ref{fig:BD} shows the schematic of a cartpole inside the simulator block. \(x\) denotes the position of the cart, and \(\theta\) is the angular displacement of the pole about the vertically upward unstable equilibrium position (UEP) ($\theta = 0^c$). The control objective is to balance the pole for small angle perturbations about the UEP, by applying a positive or negative force $F$ (two control inputs, hence $\kappa=2$) on the cart, to move it to the right or left, respectively. In this work, the \emph{CartPole-v1} simulator from OpenAI Gym \cite{b13} has been used.

Each simulation episode ends when $|\theta|>0.418^c$ or $|x|>2.4m$ (control fails) or if the cartpole is balanced for $250$ episodes (success) \cite{b10}. A constant reward of $1$ is provided for all simulation steps.
The algorithm converges when the total reward is $\ge 200$ for at least $20$ consecutive episodes \cite{b10}. Consequently, each episode can be terminated after the total reward condition is met. 
In the proposed algorithm, a reward of $-10$ is assigned when the pole falls over to assign a higher penalty for failure and enable better convergence. Figure \ref{fig:DisFlow} illustrates the discretization of the cartpole states into $10$ bins.
The systematic implementation of the synaptic Q-learning algorithm is presented in detail in a subsequent section.

\section{Algorithm} \label{algorithm}
The proposed algorithm uses neurons and synapses to store and process RL data for the controller as a directed graph.
This is discussed in detail in the next sections.

\subsection{Python Implementation: Synaptic Q-Learning Algorithm} 
As the state $s$ is received by the controller, comprising the neural network, in the forward pass, a neuron $N_i$ corresponding to state $s$ gets spawned from class $\mathcal{N}$. $N_i$ possesses synapses $S_{ij} \in \mathcal{J}$, and gates $G_{ij} \in \mathcal{G}$, representing the possible actions $a \in \mathcal{A}$ that can be taken from that state. This neuron activates its synapses by emitting a spike train with unit frequency 
$\delta(1)$. Each activated synapse emits a spike train ($\delta(f_j)$) with a frequency ($f_j$) proportional to the Q-value ($f_j = c \times Q_{ij}(s, a)$) stored in the synapse. A normally closed gate $G_{ij_{max}}$ at the synaptic output channel opens up only for the synapse ($S_{ij_{max}}$) with the maximum $f$, selecting the corresponding action $a$. During training, the value function $V_{i+1}(s')$ for the next neuron and the reward $r$ received from the environment are used to update $Q_{ij_{max}}(s, a)$  and the value function $V_i(s)$ for $N_i$ following (\ref{BellmanAlt}) and (\ref{value}), respectively.
\begin{algorithm}
 \small
    \caption{$\epsilon$-Greedy Synaptic Q-Learning}
    \label{Algo:1}
    \begin{algorithmic}[1]
       \STATE Initialise state list $\mathcal{S} = \{\}$, neuron population $\mathcal{P} = \{\}$
       \STATE Initialize $\gamma$, $\tau = -\frac{1}{ln(\gamma)}$,  $\alpha$, and spike function $\delta(f)$ 
       \STATE Initialise action space $\mathcal{A} = \{a_{j}\}$, $\forall j\in\{1,2, \dots, \kappa \}$ 
       \STATE Initialize classes $\mathcal{N}(V(s),\overline{c},\overline{f})$, $\mathcal{J}(Q(s,a),\overline{c},\overline{f},\tau)$, $\mathcal{G}(\overline{c})$ 
       \STATE $\mathcal{G}(\overline{c})$ $\gets$ close
       \WHILE{not converged}
           \STATE Start in state $\textbf{s} \in \mathcal{S}$
           \STATE Initialize neuron $N_i \in \mathcal{N}$
           \STATE Initialize synapses $S_{ij} \in \mathcal{J}$ and gates $G_{ij} \in \mathcal{G}$
           \STATE Append $\textbf{s}$ to $\mathcal{S}$ 
           \STATE Append $N_{i}$ to $\mathcal{P}$
           \STATE $j = 0$
           \WHILE{episode not terminated} 
                    \WHILE{$j \leq \kappa$}
                        \STATE $S_{ij}$ gets $\delta(1)$ from $N_{i}$,  
                        \STATE $G_{ij}$ gets $\delta(f_{j})$ where $f_{j} = c \times Q_{ij}(\textbf{s},\textbf{a})$
                        \STATE $j = j + 1$
                    \ENDWHILE
                    \STATE $j_{max}$ $\gets$ $\argmax_{j}(f_{j})$ 
                    
                    \STATE $G_{ij_{max}}$ $\gets$ open
                    \STATE Select action $\textbf{a}_{j_{max}}$
                    \STATE Send $\textbf{a}_{j_{max}}$ to simulator
                    \STATE Receive $r$,$\textbf{s}^{\prime}$ from simulator
                    \IF{$\textbf{s}^{\prime} \notin \mathcal{S}$}
                       \STATE Initialize $N_{i+1} \in \mathcal{N}$,  $S_{(i+1)j} \in \mathcal{J}$, and $G_{(i+1)j} \in \mathcal{G}$, corresponding to $\textbf{s}^{\prime}$,  $\forall j\in \{1,2, \dots, \kappa\}$  
                       \STATE Append $\textbf{s}^{\prime}$ to $\mathcal{S}$ 
                       \STATE Append $N_{i+1}$ to $\mathcal{P}$
                    \ELSE 
                        \STATE Choose $N_{i+1} \in \mathcal{P}$ corresponding to $\textbf{s}^{\prime}$
                    \ENDIF
                    \STATE Connect $S_{ij_{max}}$ to $N_{i+1}$ via $G_{ij_{max}}$
                    \STATE $Q_{ij_{max}}(\textbf{s},\textbf{a}) \gets Q_{ij_{max}}(\textbf{s},\textbf{a}) + \alpha (-Q_{ij_{max}}(\textbf{s},\textbf{a}) + r + \gamma V_{i+1}(\textbf{s}^{\prime}))$ 
                    \STATE $V_{i}(\textbf{s}) \gets \max_{j} (Q_{ij}(\textbf{s},\textbf{a}))$ 
                    \STATE $\textbf{s} \gets \textbf{s}'$
                    \STATE $N_{i} \gets N_{i+1}$
                    \STATE $G_{ij}$ $\gets$ close
           \ENDWHILE
       \ENDWHILE
    \STATE \textbf{end}
    \end{algorithmic}
\end{algorithm}
\setlength{\textfloatsep}{0pt}

\begin{algorithm}
 \small
    \caption{Bellman Memory Units in Nengo}
    \label{Algo:2}
    \begin{algorithmic}[1]
       \STATE Initialise state list $\mathcal{S} = \{\}$, ensemble population $\mathcal{P} = \{\}$, action vector $\mathcal{A}=\{a_{jk}\}$, where $j\in \{0,1,2,...,b\}$ and $k\in \{0,1,2,...,d\}$
       \STATE Initialize learning rate  $\alpha$, bin size $b$, dimensions $d$, a and unit step node $u(t)$ 
       \WHILE{not converged}
           \STATE Initialize state $\textbf{s}$. Append $\textbf{s}$ to $\mathcal{S}$
           \STATE Initialize nengo ensemble $N_i$ with a label $\textbf{s}$, with $b$ neurons, $d$ dimensions, randomly chosen encoders $\{e_{(i)_{jk}}\}$  $\forall j\in \{1,2, \dots, b\}$ and   $k\in \{1,2, \dots, d\}$
           \STATE Initialize $Q$ value to each neuron in the ensemble to be equal to the encoder for that neuron.
           \STATE Append $N_{i}$ to $\mathcal{P}$
           \STATE Connect $u(t)$ node to $N_{i}$
           \WHILE{nengo simulator runs}
               \WHILE{episode not terminated} 
                    \STATE Extract activity $a_{ijk}\in \mathcal{A}$ from $N_i$ $\forall j\in\{1,2, \dots, b\}$ along each dimension $k$
                    \STATE $j_{max_k}$ $\gets$ $\argmax_{j}(a_{ijk})$ $\forall k$ $\in \{1,2, \dots , d\}$
                    
                    \STATE Select action $\mathbf{a}_{{ij_{max}}_{(d\times 1)}}$
                    \STATE Send $\mathbf{a}_{{ij_{max}}_{(d\times 1)}}$ to simulator
                    \STATE Receive $r$,$\textbf{s}^{\prime}$ from simulator
                    \IF{$\textbf{s}^{\prime} \notin \mathcal{S}$}
                       \STATE Initialize $N_{i+1}$ with $b$ neurons and $d$ dimensions corresponding to $\textbf{s}^{\prime}$, with randomly chosen encoders $\{e_{(i+1)_{jk}}\}$  $\forall j\in \{1,2, \dots, b\}$ and $k\in \{1,2, \dots, d\}$
                       \STATE Initialize $Q$ value to each neuron in the ensemble to be equal to the encoder for that neuron.
                       \STATE Append $\textbf{s}^{\prime}$ to $\mathcal{S}$ 
                    \ELSE 
                        \STATE Choose $N_{i+1} \in \mathcal{P}$ corresponding to $s'$
                    \ENDIF
                    \STATE Define nengo connection from node $u(t)$ to $N_{i+1}$
                    \STATE Define nengo connection from $N_i$ to $N_{i+1}$
                    \STATE $Q(\textbf{s},\textbf{a})$$ \gets Q(\textbf{s},\textbf{a}) + \alpha (-Q(\textbf{s},\textbf{a}) + r + \gamma V(\textbf{s}^{\prime}))$ 
                    \STATE $V(\textbf{s}^{\prime}) \gets \max_{\textbf{a}^{\prime}}(Q(\textbf{s}^{\prime},\textbf{a}^{\prime}))$ 
                    \STATE $\textbf{e}[\textbf{a}_{i(b \times 1)_{j_{max}}}] \gets Q(\textbf{s},\textbf{a})$
                    \STATE $\textbf{s} \gets \textbf{s}^{\prime}$
                    \STATE $N_{i} \gets N_{i+1}$
               \ENDWHILE
            \ENDWHILE
       \ENDWHILE
    \STATE \textbf{end}
    \end{algorithmic}
\end{algorithm}

\begin{figure}[htb]
\centerline{\includegraphics[clip=true, trim={2.8cm 0 3.3cm 1cm},width=9cm]{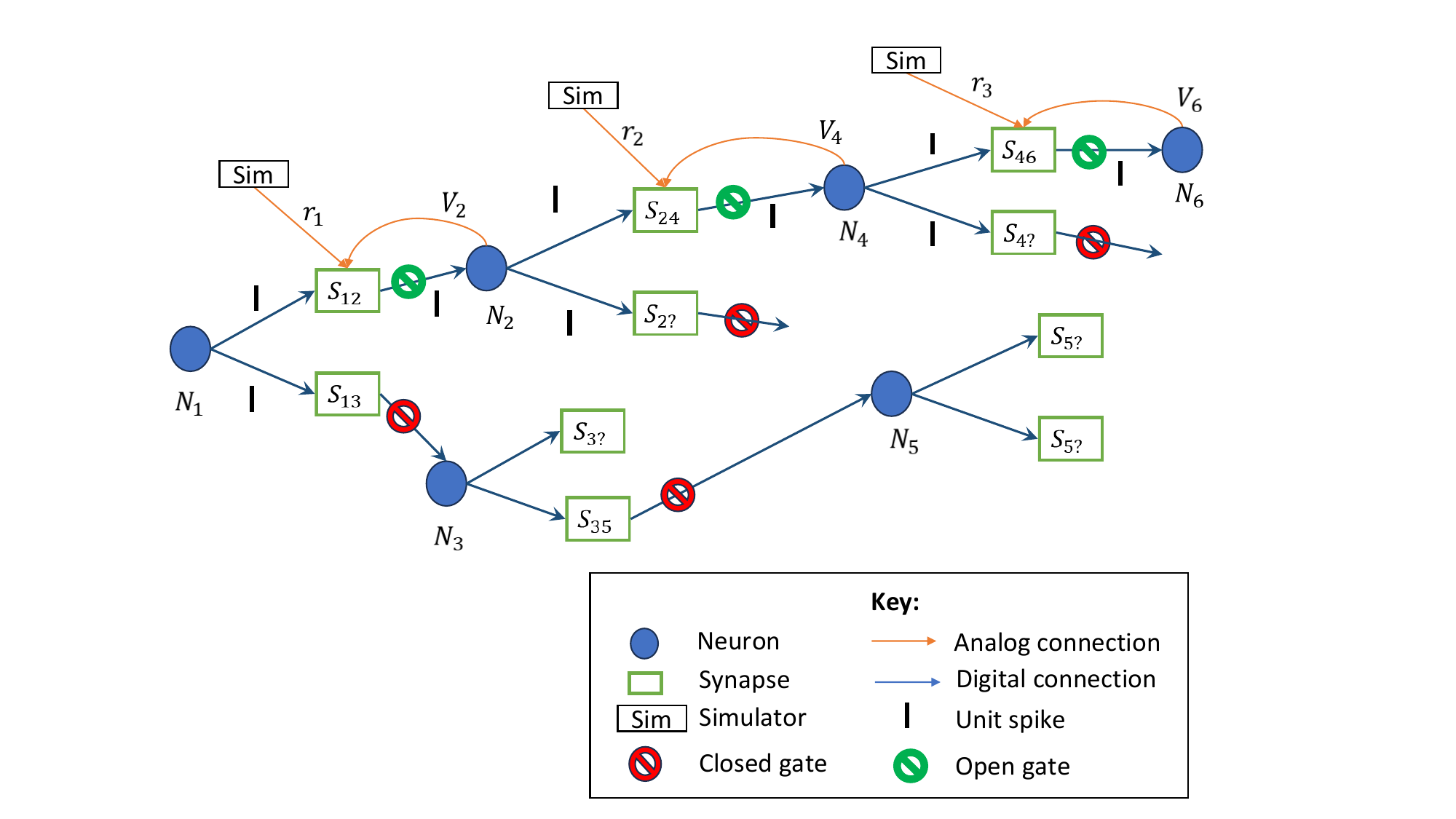}}
\caption{An example network state for training an arbitrary synaptic Q-learning model. The spike propagation is shown for a particular forward pass of the network.  $r_i$ is the reward received for the $i^{th}$ observation and $V_i$ is the magnitude of the value function for the $i^{th}$ neuron.
}
\label{fig:Flow}
\end{figure}

A list of objects used in the network and the data they store are shown in Table \ref{table:symbol}.  Blue text in the table corresponds to blue arrows in Fig. \ref{fig:Flow}, which represent non-learnable digital connections, while the orange ones represent learnable analog connections. Fan-in refers to the number of input connections to an neuron, which is an important parameter to design a neuron element on chip. 

This discounting time factor $\gamma$ can be included in a spiking neuron synapse as the temporal aspect is built-in in the discrete convolution shown in (\ref{syncurrent}).
Considering, discrete simulation time intervals of $1$s, the synaptic update occurs at $t=1$ and $V(s^{\prime})$ of the next neuron is taken from the previous forward pass at $t=0$ for performing such update. A simplified version of the synaptic operation can be formed (\ref{simpl_syncurrent}).
\vspace*{-0.2cm}
\begin{align}
I_{syn}(t) &= I_{in}(0) e^{-(\frac{1}{\tau})} \nonumber \\
\gamma &\sim e^{-(\frac{1}{\tau})}
\label{simpl_syncurrent}
\end{align}
where, $I_{in}(0)$ is analogous to $V(s^{\prime})$ in the Bellman equation.



A higher value of the $\tau$ means a higher $\gamma$ and consequently, more time till which future rewards are collected. 
The pseudo-code for $\epsilon$-greedy synaptic Q-learning algorithm is presented in Algorithm \ref{Algo:1}.

In summary, synaptic Q-learning has the following salient features
\begin{itemize}
 \item Evolution of network topology, both number of neurons and connections, with training.
 \item Forward pass (digital using spikes) for action selection and recurrent connections (analog) for training using the Bellman equation at the synapse \cite{b4,b5,b6}.
 \item Discounting factor $\gamma$ in-built in the synaptic time-constant, and action-selection gates ($\mathcal{G}$) at the synaptic output.
 
\end{itemize}

\begin{figure*}[htb]
\centering
\subfloat[1 episode]{
  \includegraphics[clip=true, trim={0cm 0cm 0 0cm},width=5cm]{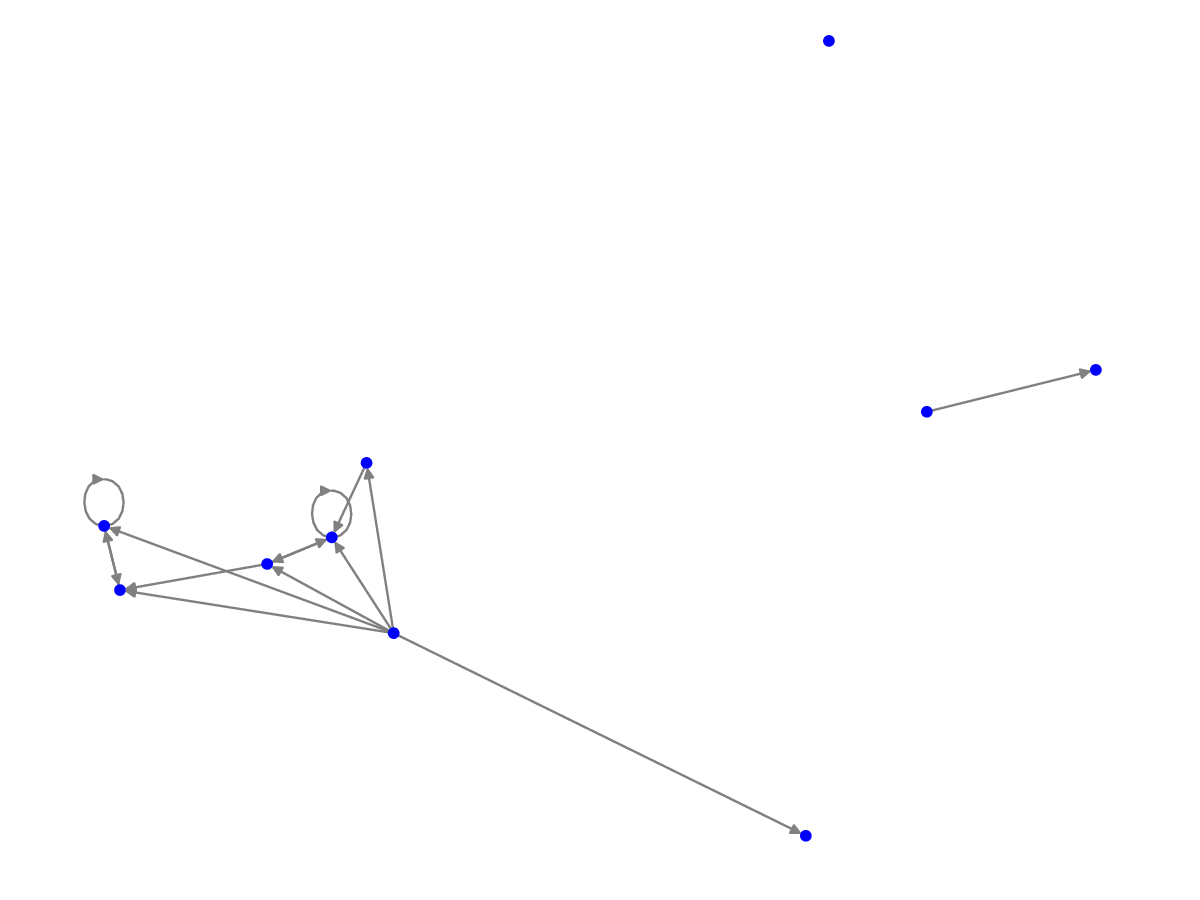}
}
\subfloat[10 episodes]{
  \includegraphics[clip=true, trim={0cm 0cm 0 0cm},width=5cm]{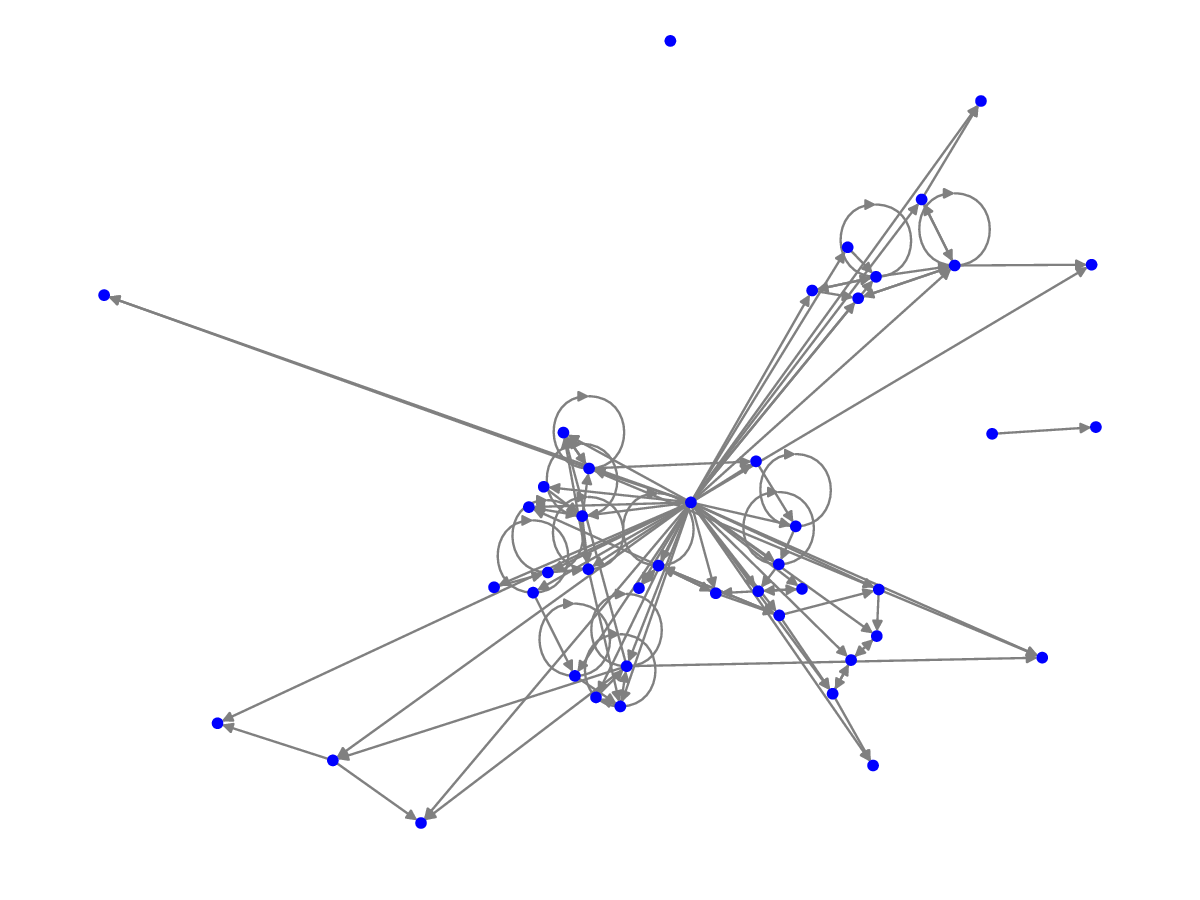}
}
\subfloat[20 episodes]{
  \includegraphics[clip=true, trim={0cm 0cm 0 0cm},width=5cm]{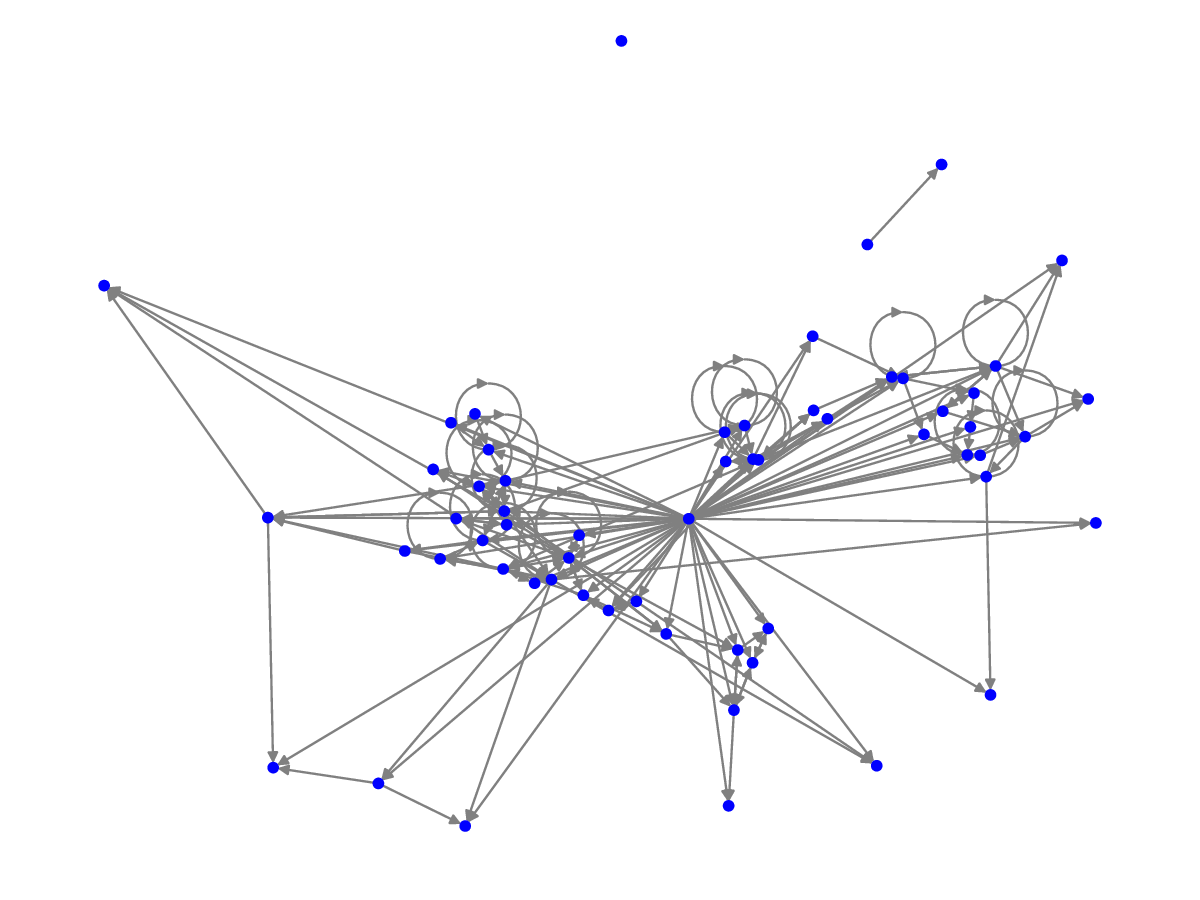}
}
\hspace{0mm}
\subfloat[30 episodes]{
  \includegraphics[clip=true, trim={0cm 0cm 0 0cm},width=5cm]{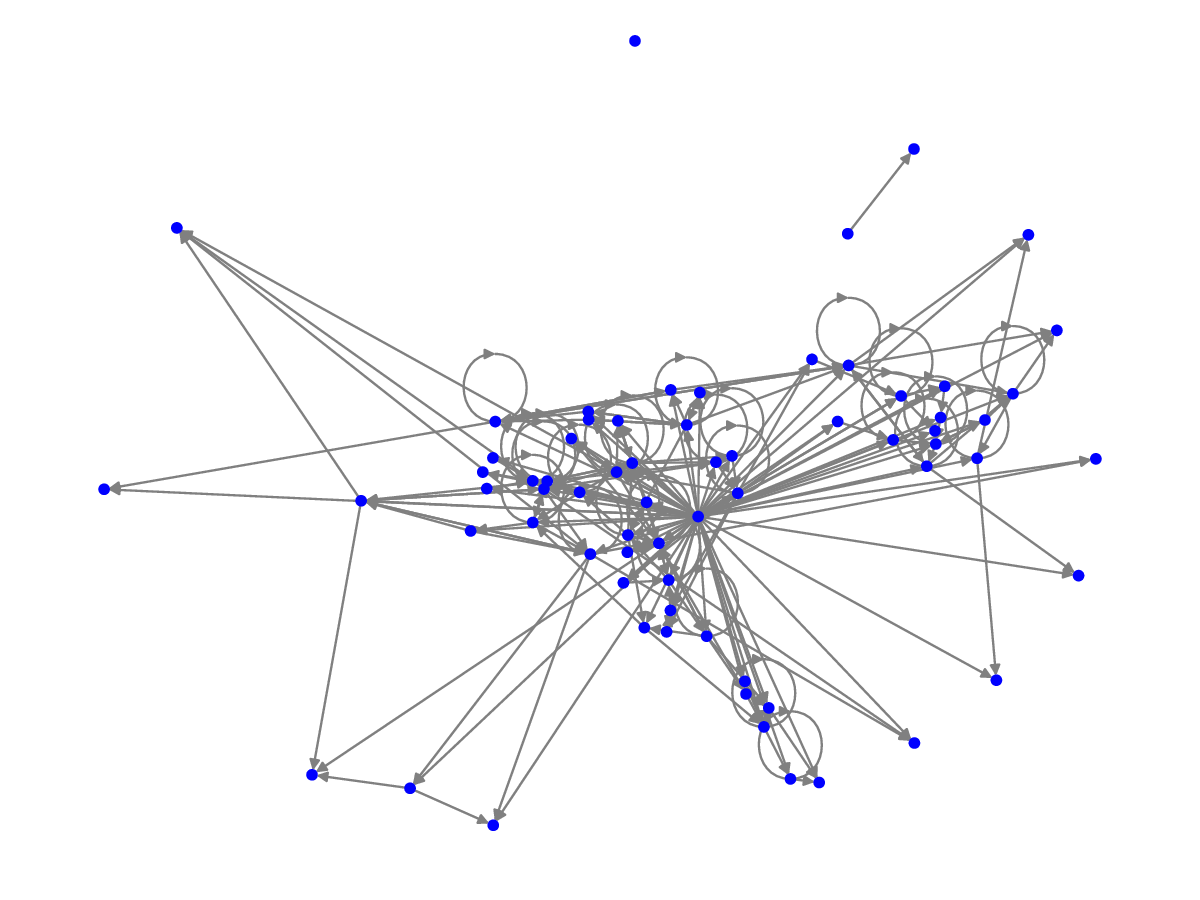}
}
\subfloat[40 episodes]{
  \includegraphics[clip=true, trim={0cm 0cm 0 0cm},width=5cm]{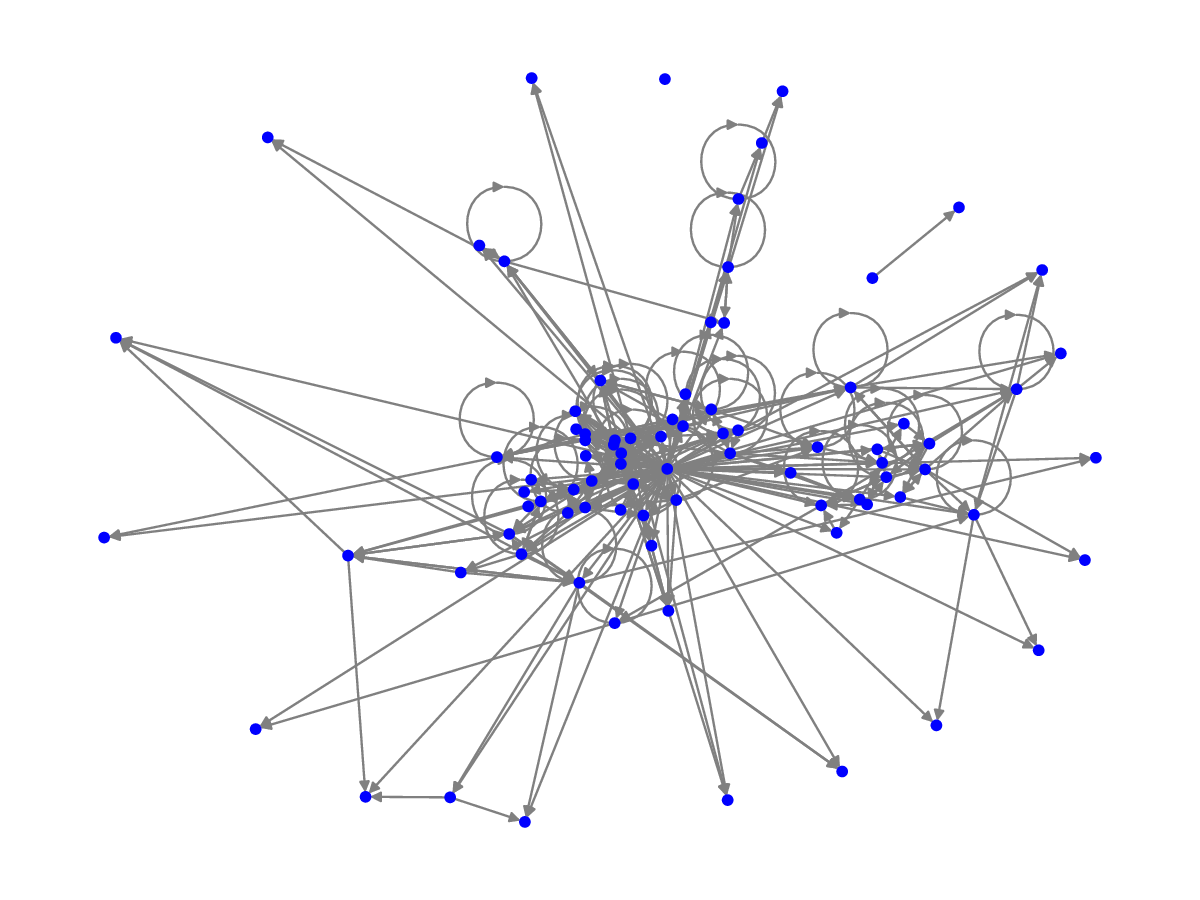}
}
\subfloat[50 episodes]{
  \includegraphics[clip=true, trim={0cm 0cm 0 0cm},width=5cm]{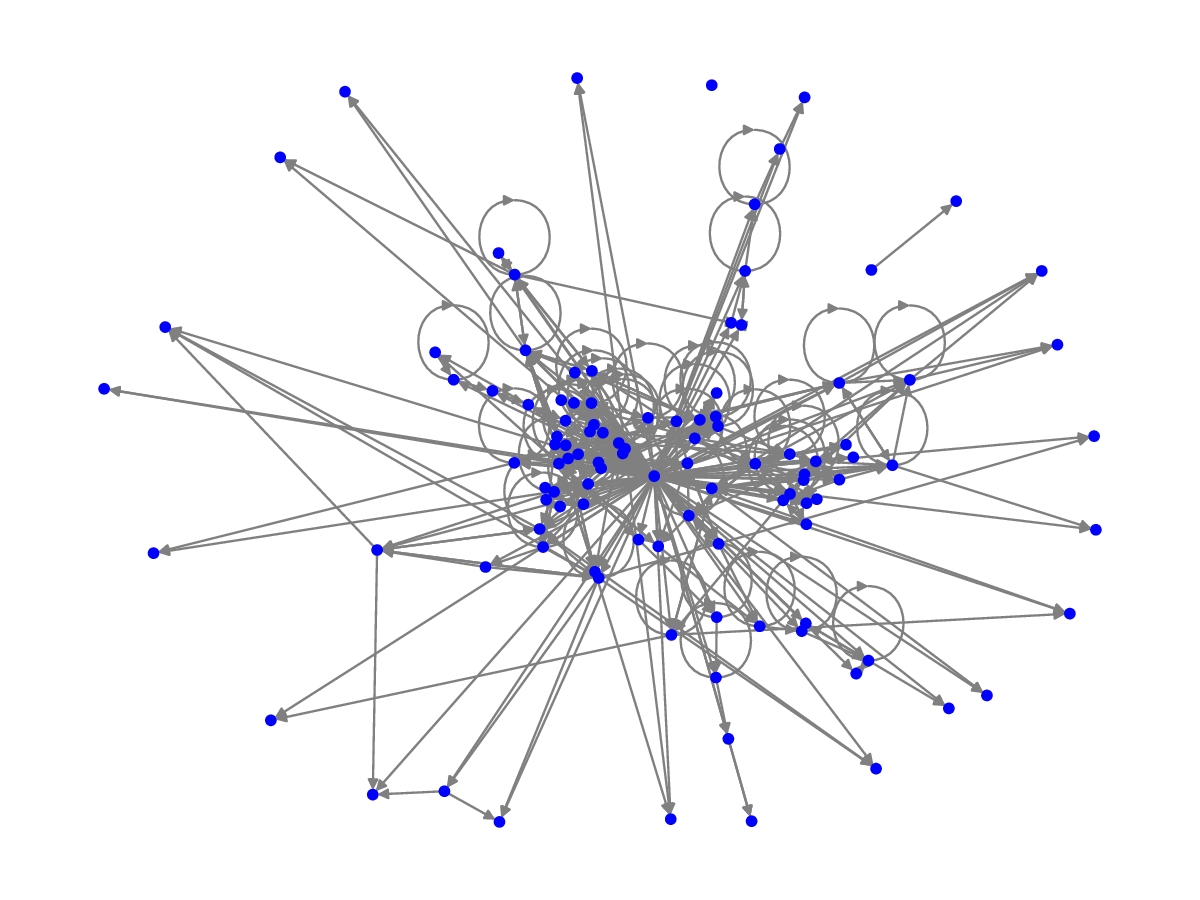}
}
\hspace{0mm}
\subfloat[60 episodes]{
  \includegraphics[clip=true, trim={0cm 0cm 0 0cm},width=5cm]{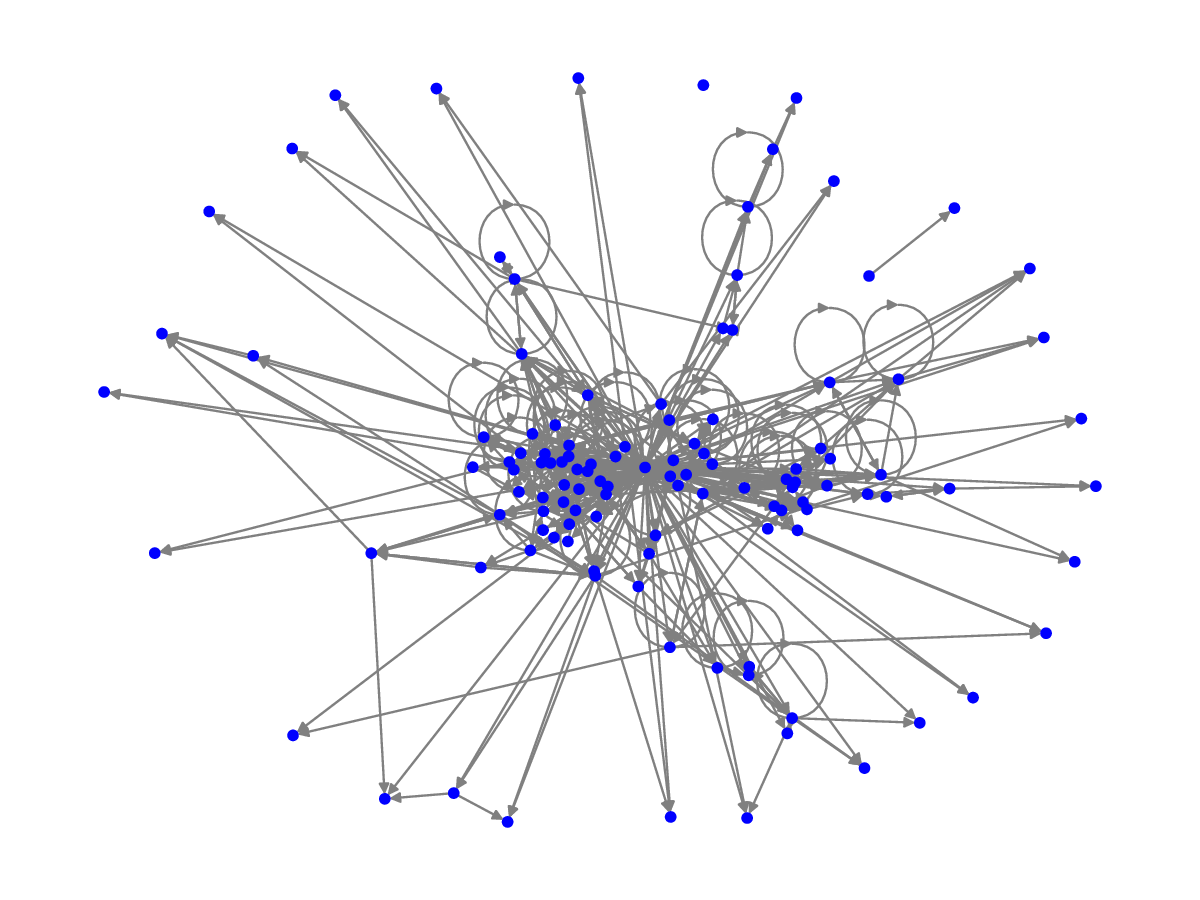}
}
\subfloat[70 episodes]{
  \includegraphics[clip=true, trim={0cm 0cm 0 0cm},width=5cm]{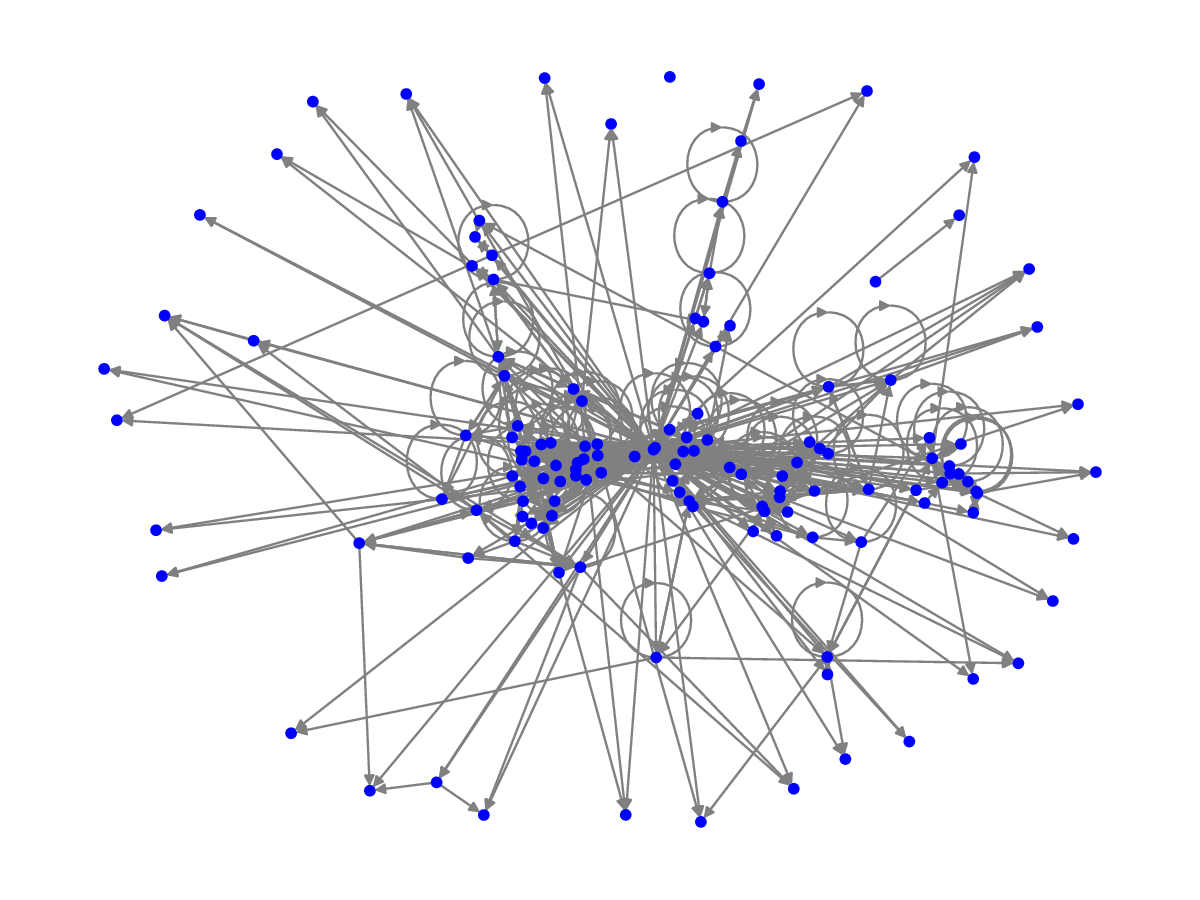}
}
\subfloat[80 episodes]{
  \includegraphics[clip=true, trim={0cm 0cm 0 0cm},width=5cm]{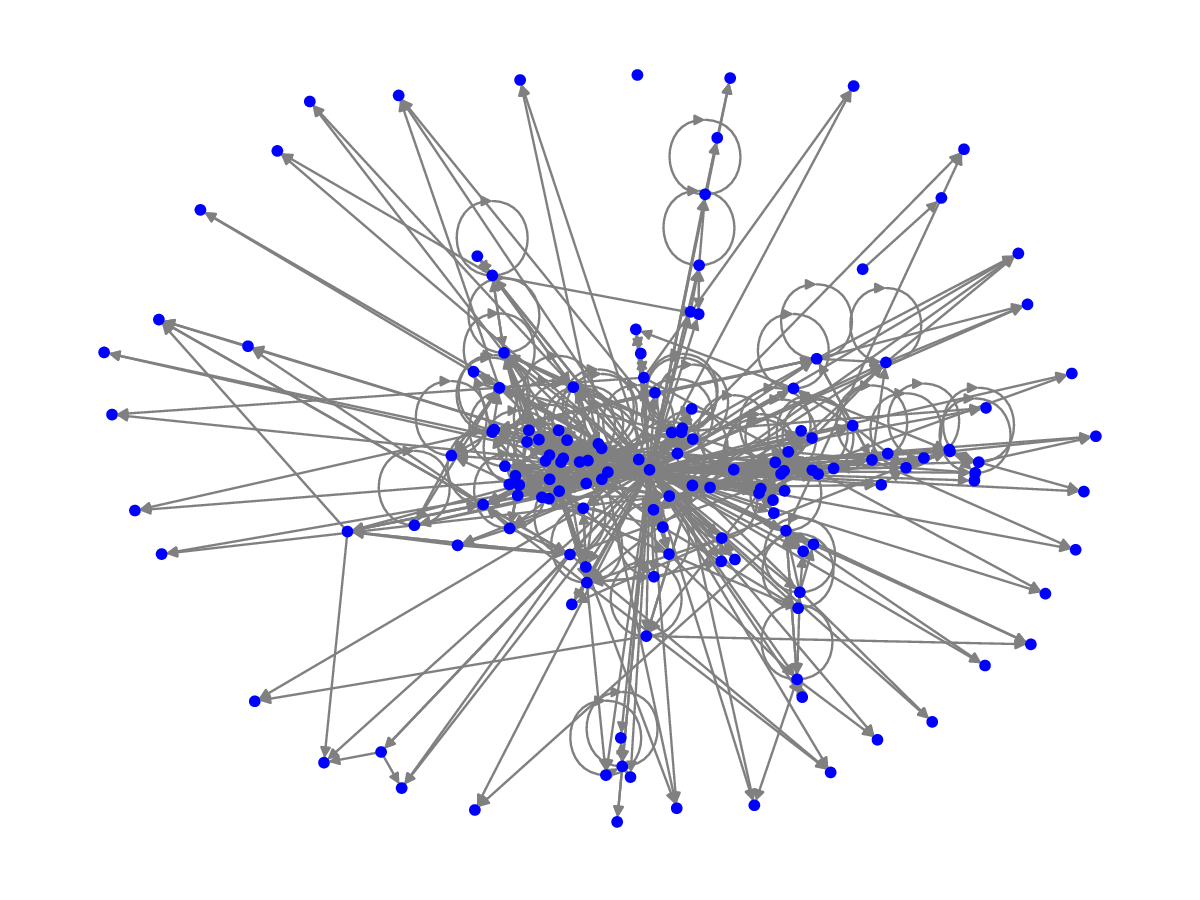}
}
\hspace{0mm}
\caption{Figure showing the evolution of network architecture in Nengo. The blue dots symbolize nodes and ensembles and the gray lines symbolize the connections between them.}
\label{fig:nengo_evolution}
\end{figure*}

\begin{figure*}[htb]
\centering
\subfloat[1 episode]{
  \includegraphics[clip=true, trim={0cm 0cm 0 0cm},width=5cm]{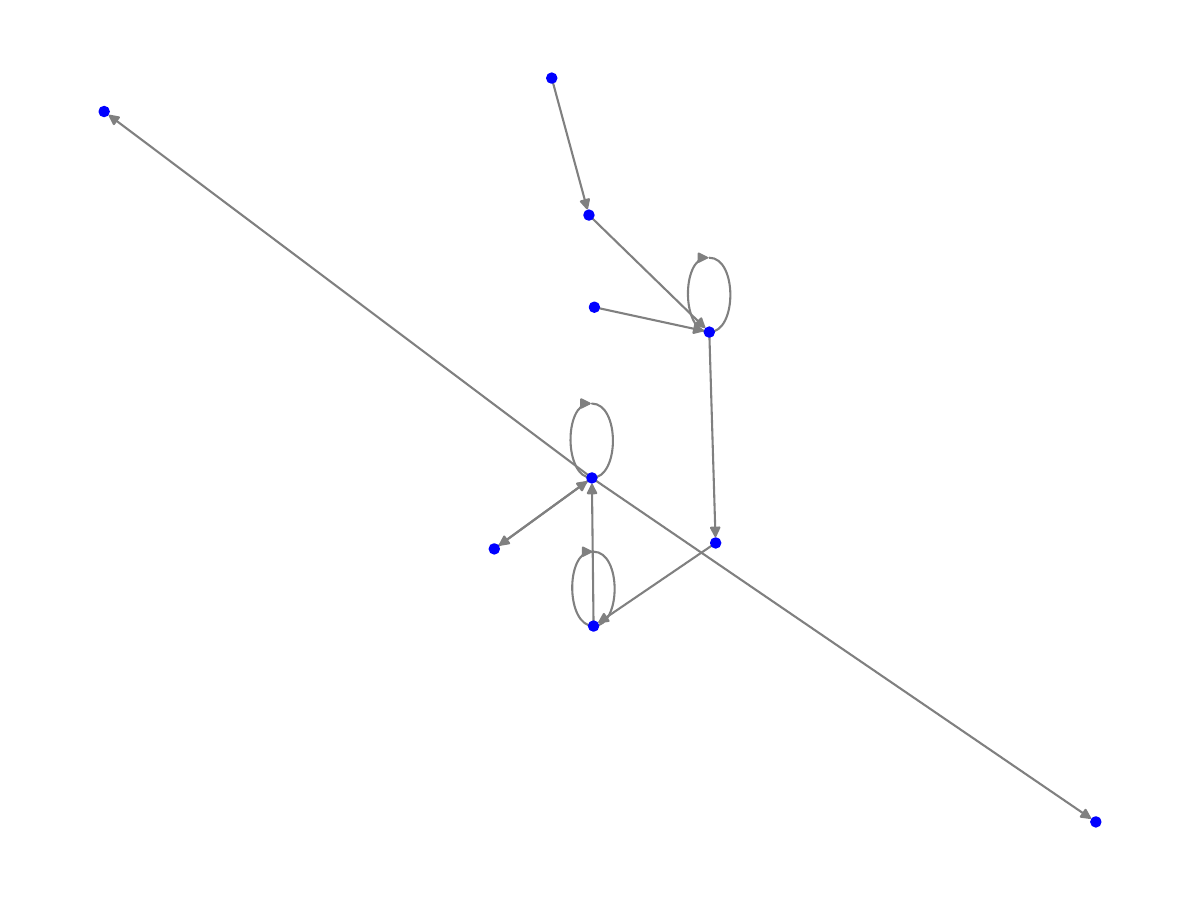}
}
\subfloat[10 episodes]{
  \includegraphics[clip=true, trim={0cm 0cm 0 0cm},width=5cm]{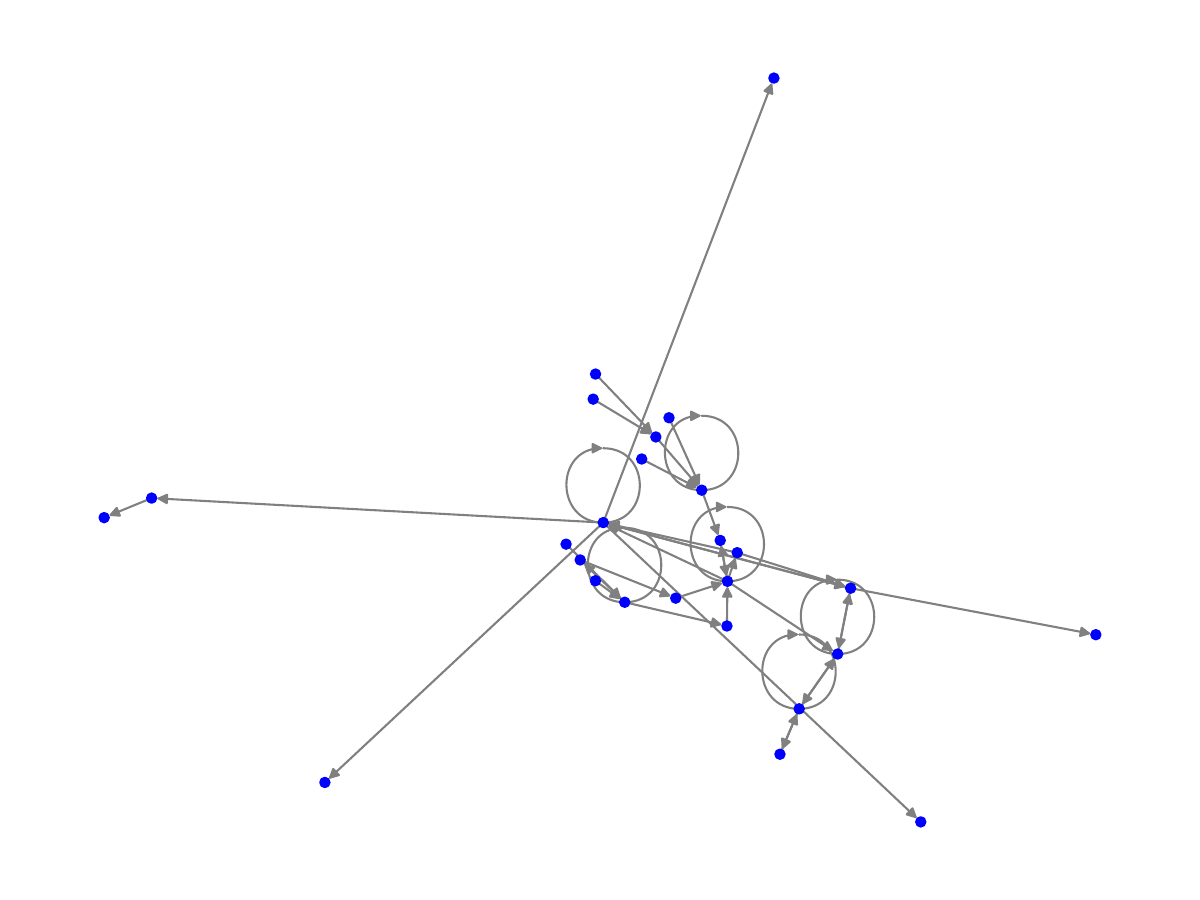}
}
\subfloat[20 episodes]{
  \includegraphics[clip=true, trim={0cm 0cm 0 0cm},width=5cm]{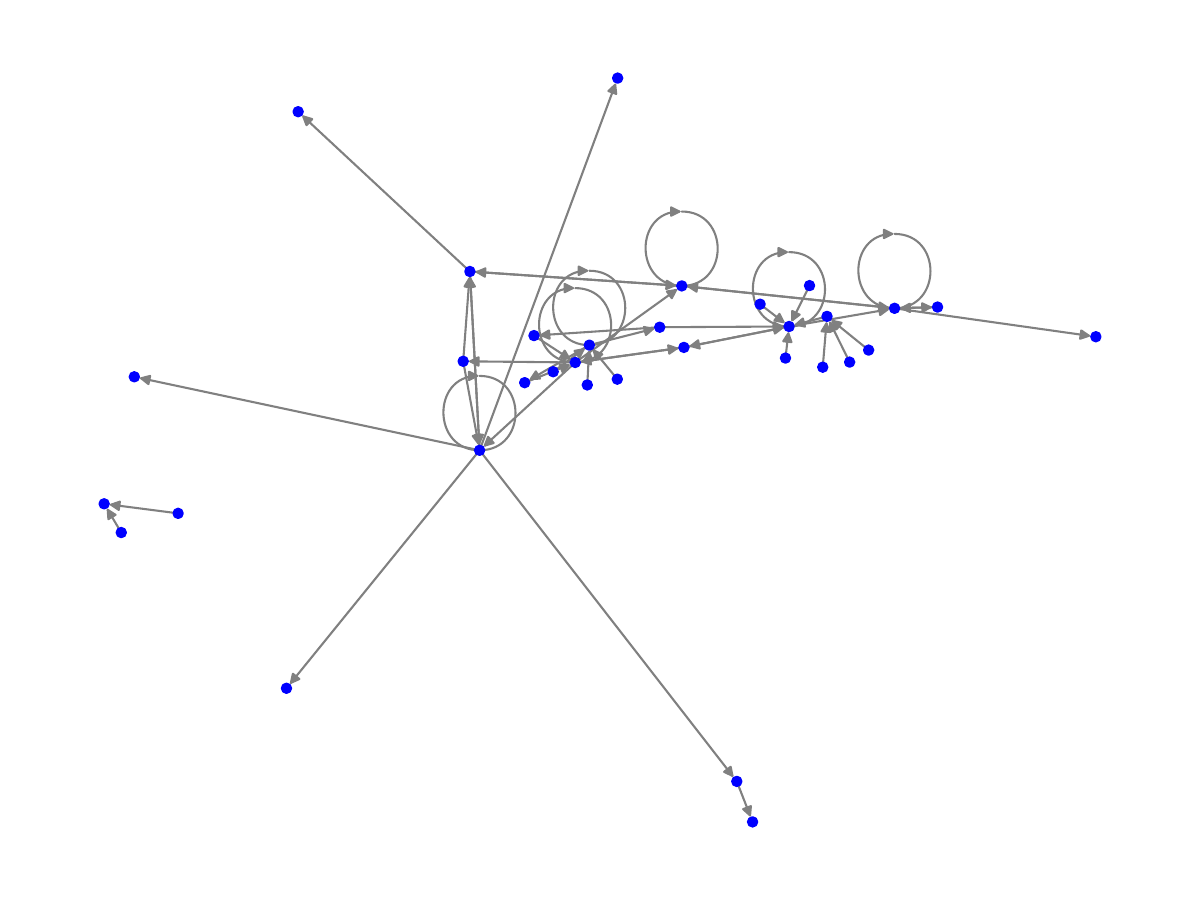}
}
\hspace{0mm}
\subfloat[30 episodes]{
  \includegraphics[clip=true, trim={0cm 0cm 0 0cm},width=5cm]{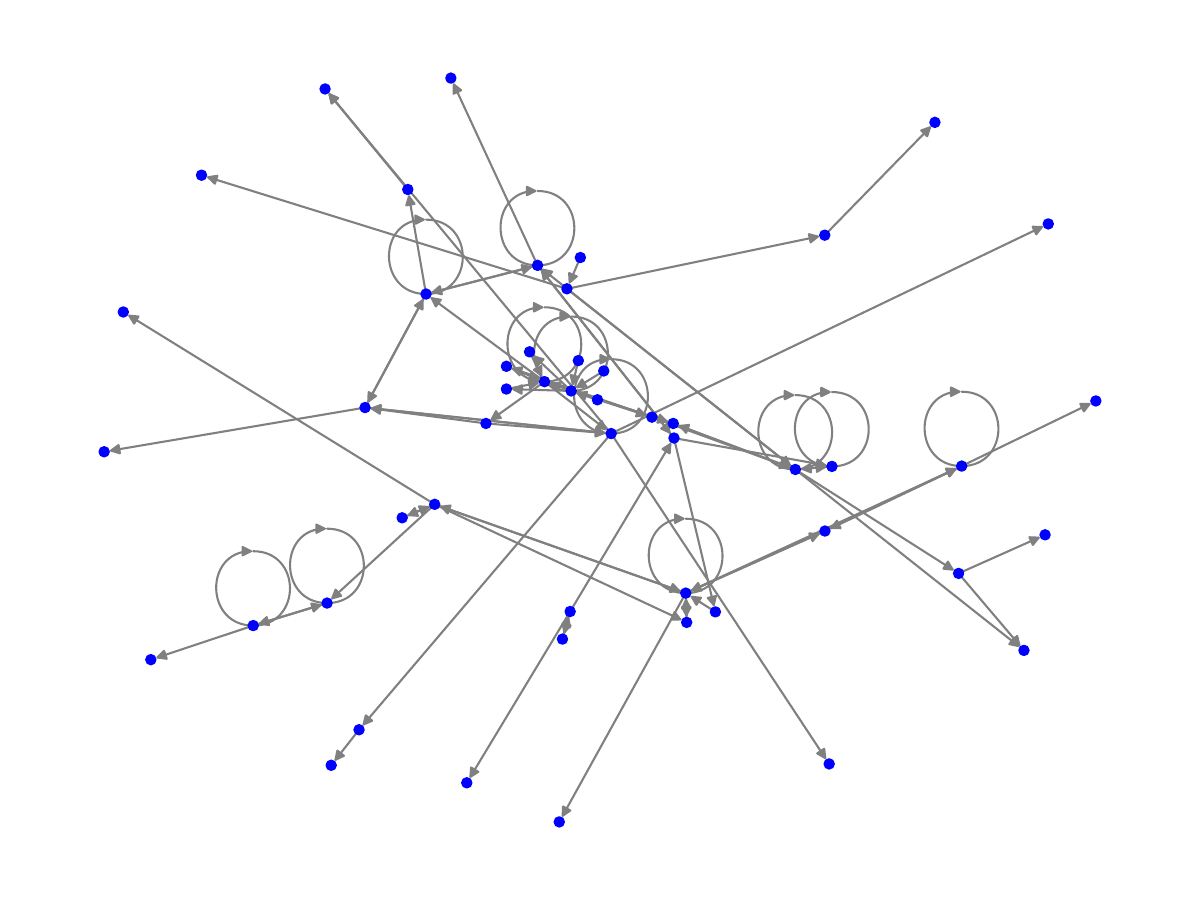}
}
\subfloat[40 episodes]{
  \includegraphics[clip=true, trim={0cm 0cm 0 0cm},width=5cm]{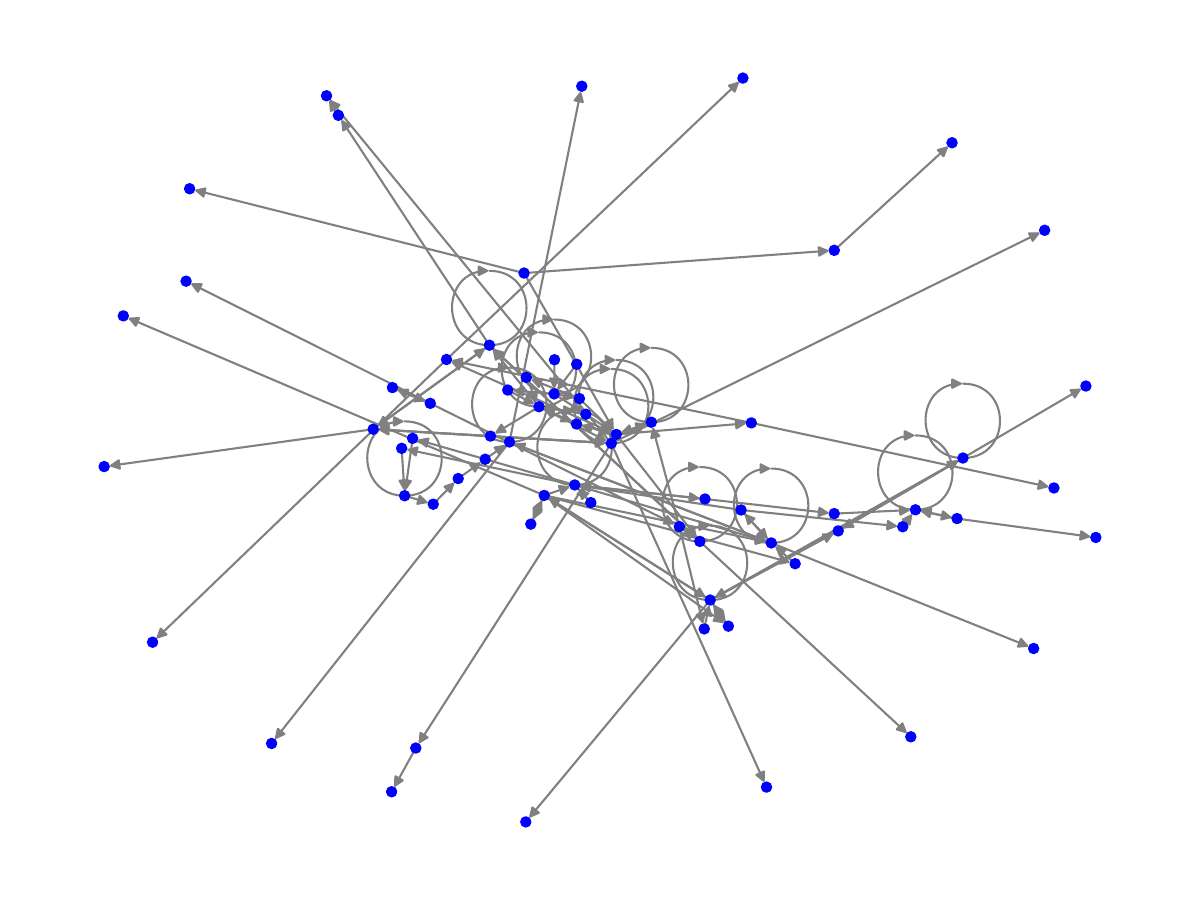}
}
\subfloat[50 episodes]{
  \includegraphics[clip=true, trim={0cm 0cm 0 0cm},width=5cm]{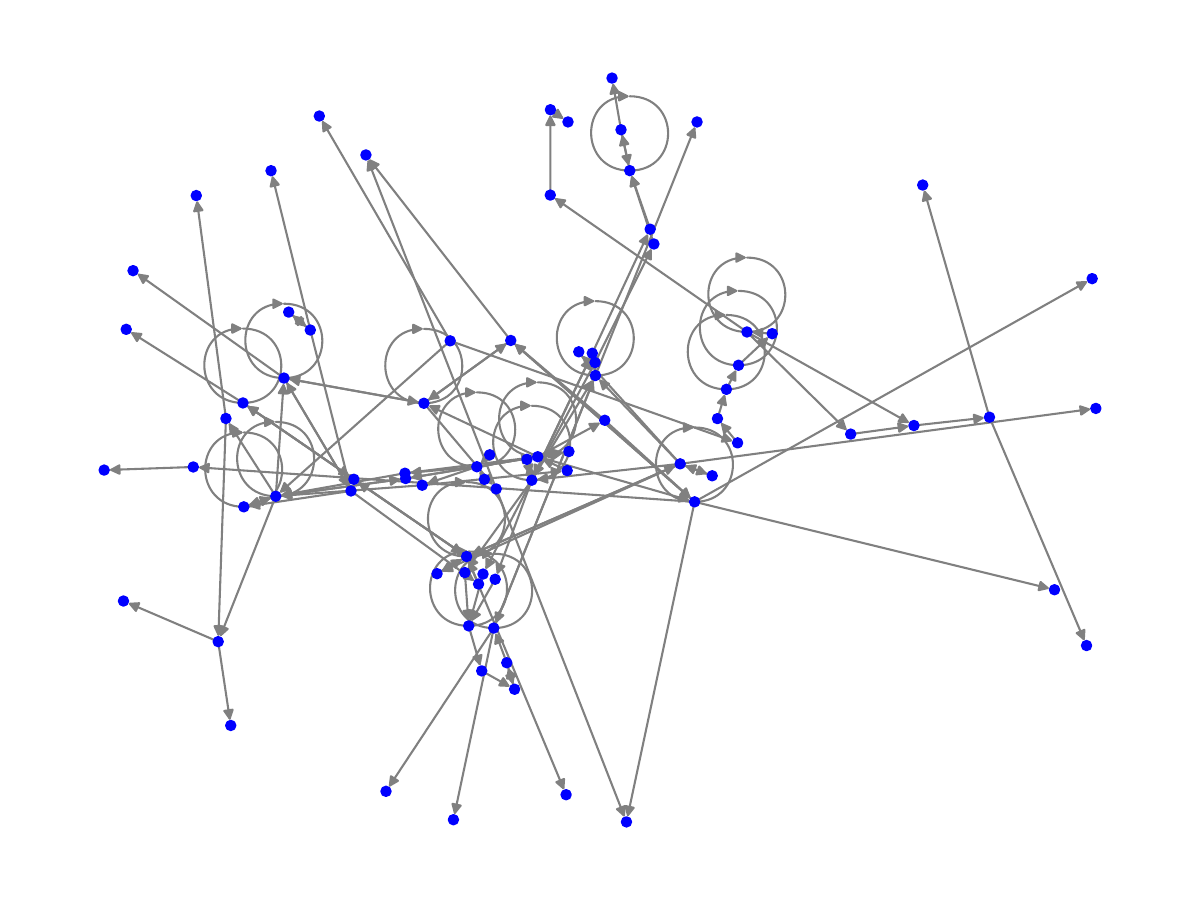}
}
\hspace{0mm}
\subfloat[60 episodes]{
  \includegraphics[clip=true, trim={0cm 0cm 0 0cm},width=5cm]{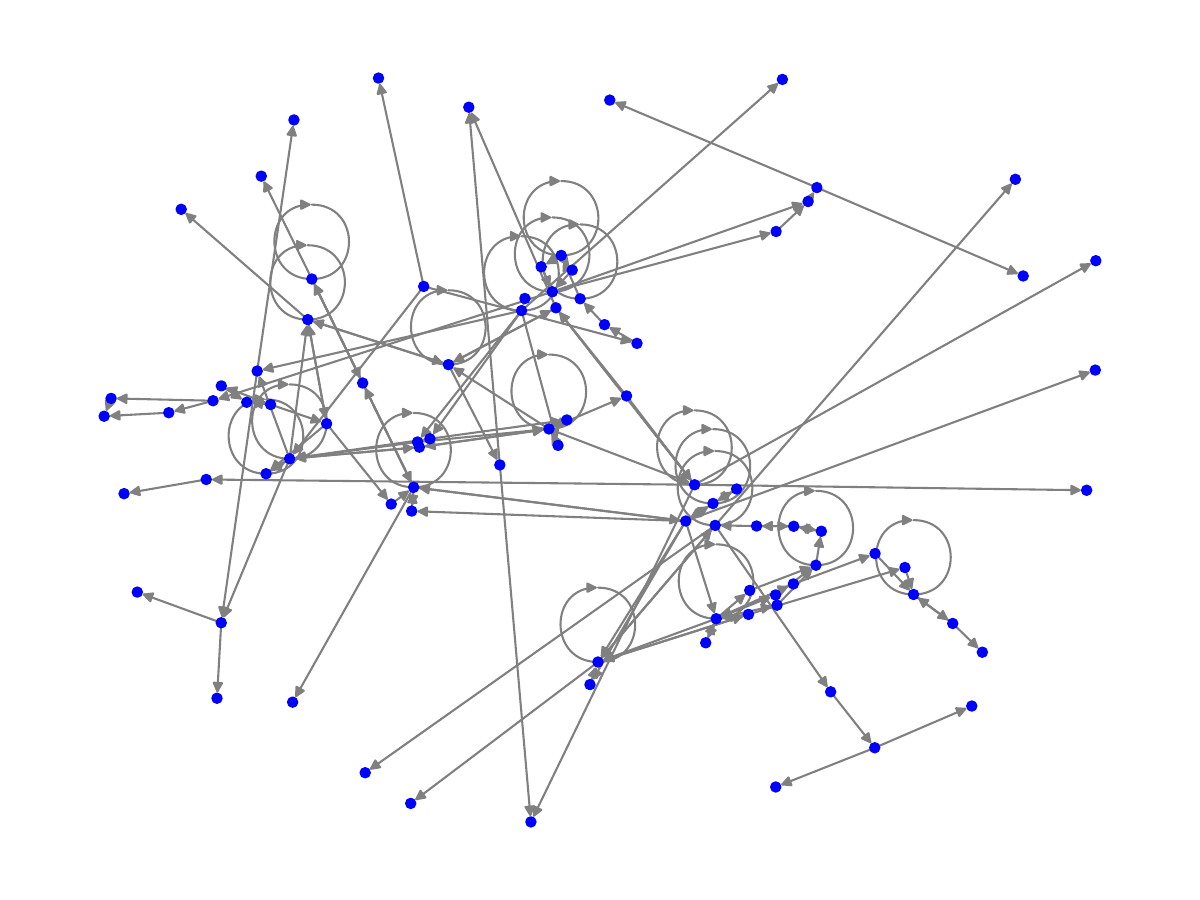}
}
\subfloat[70 episodes]{
  \includegraphics[clip=true, trim={0cm 0cm 0 0cm},width=5cm]{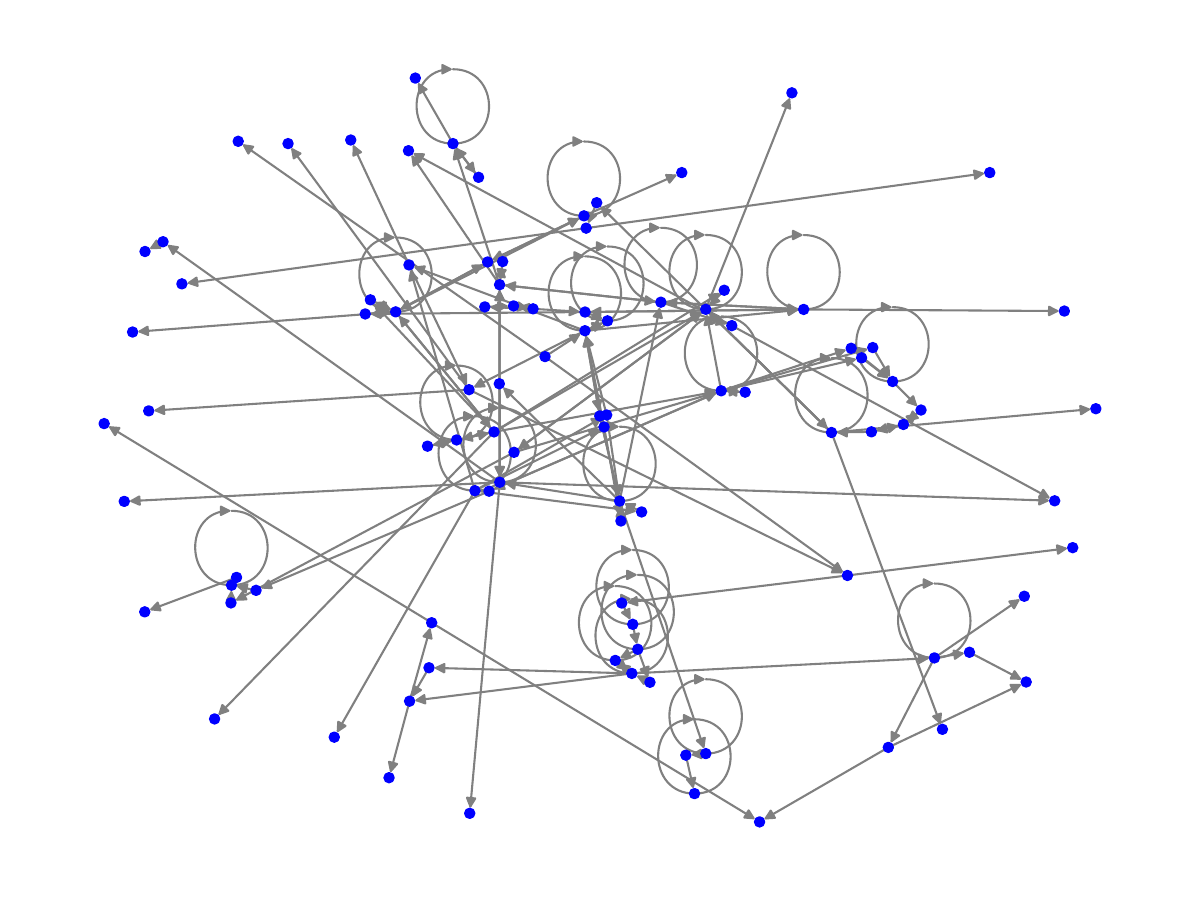}
}
\subfloat[80 episodes]{
  \includegraphics[clip=true, trim={0cm 0cm 0 0cm},width=5cm]{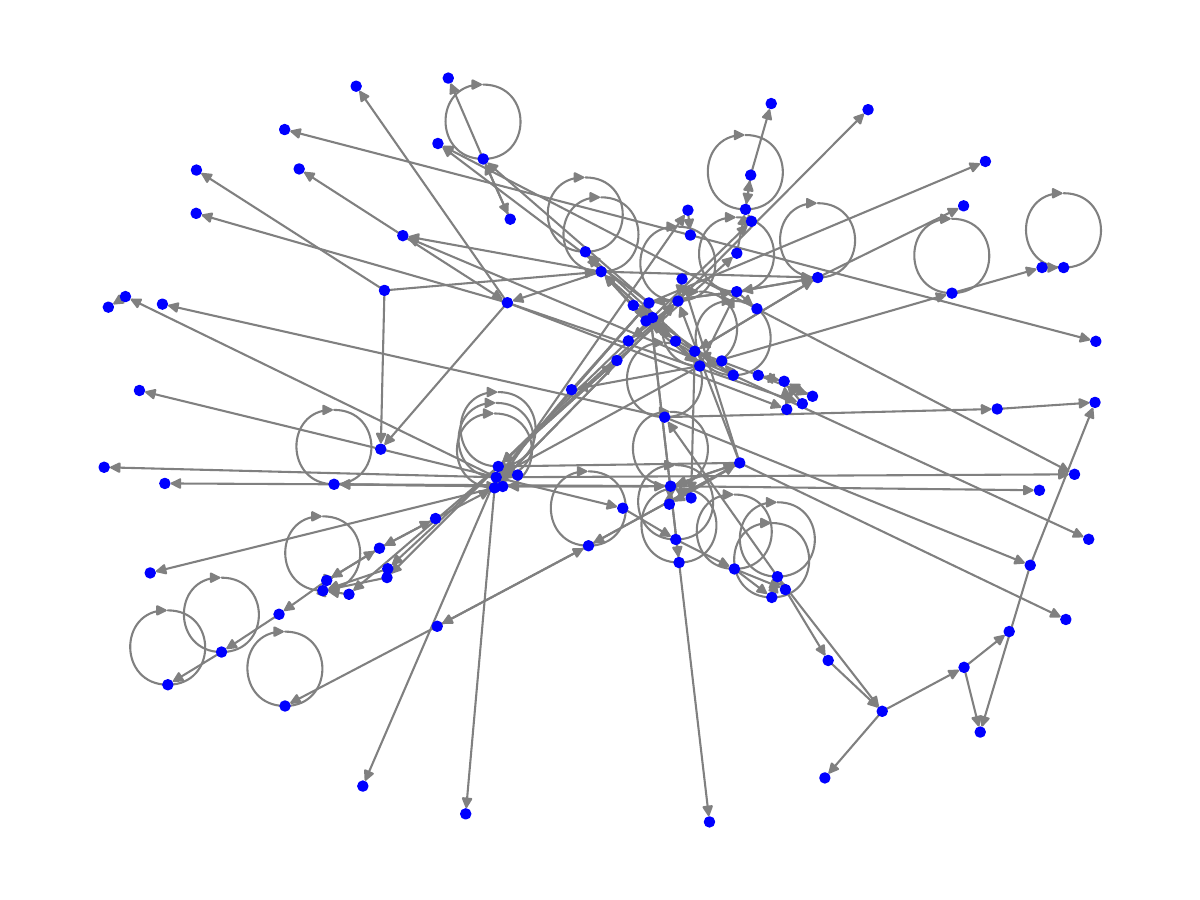}
}
\hspace{0mm}
\caption{Figure showing the evolution of network architecture on Loihi board using online learning. The blue dots symbolize nodes and ensembles and the gray lines symbolize the connections between them.}
\vspace{3mm}
\label{fig:loihi_evolution}
\end{figure*}

\subsection{Implementation in the Nengo NEF: Bellman Memory Units}
To validate this algorithm in an SNN platform, the NEF is chosen, where ensembles of spiking neurons are dynamically created whenever a new state of the system is encountered. The dimension of the ensemble $d$ is chosen to be equal to the dimension of the action space of the system. The number of neurons in the ensemble is equal to the number of bins $b$ into which each control action value is discretized. Each neuron denotes a unique value of control action sampled from the vector $\mathcal{A}$. The Q values are initialized randomly with the values of the neuron's encoders. At every timestep, a distinct ensemble corresponding to the discretized value of the system state is either spawned or chosen from the group of already spawned ensembles in the population vector $\mathcal{P}$. The neuron with the highest activity is selected and the corresponding action value $\mathbf{a}_{{ij_{max}}_{(d\times 1)}}$ is sent to the simulator. The reward feedback obtained from the system, and the Q values from the ensemble corresponding to the next state are plugged into the Bellman equation (see \ref{BellmanAlt} and \ref{value}). This generates the updated Q value which is then stored in the neuron's encoder.

The activity (firing rate) of a neuron $j$ in the $i^{th}$ ensemble having $d$ dimensions \cite{b14} is given by Equation \ref{activity}.

\begin{equation}
\mathbf{\mathcal{a}}_{ij(d\times 1)}(t) = G[\mathbf{J}_{i(d\times 1)}(t)] 
\label{activity}
\end{equation}
\begin{equation}
\mathbf{J}_{i(d\times 1)}(t) = \alpha_{ij}<\mathbf{e}_{ij(d\times 1)},\mathbf{x}_{i(d\times 1)}(t)> + \mathbf{J}^{bias}_{i(d\times 1)}
\label{activity}
\end{equation}

where, $\mathbf{J}_{i(d\times 1)}(t)$ is the input current vector to the $j^{th}$ neuron in the $i^{th}$ ensemble, $G$ is the spiking neural non-linearity function illustrated by the tuning curves, $\mathbf{J}^{bias}_{i(d\times 1)}$ is the bias current of the corresponding neuron, $\mathbf{x}_{i(d\times 1)}(t)$ is the input vector to the ensemble at time $t$, $\mathbf{e}_{ij(d\times 1)}$ is the encoder vector for the $j^{th}$ neuron, and $\alpha_{ij}$ is the gain for the same. $<.,.>$ is the inner product between two vectors. For the LIF neurons used in this work, the activity of a neuron is proportional to the input current, which in turn is proportional to the encoder (since every neuron is provided with a constant unit step input). Hence the neuron with the maximum activity can be safely approximated to the neuron with the maximum encoder value. The steps are outlined in Algorithm \ref{Algo:2}.

\begin{figure*}[htb]
\centering
\subfloat[]{
  \includegraphics[clip=true, trim={2cm 2cm 0 1cm},width=5cm]{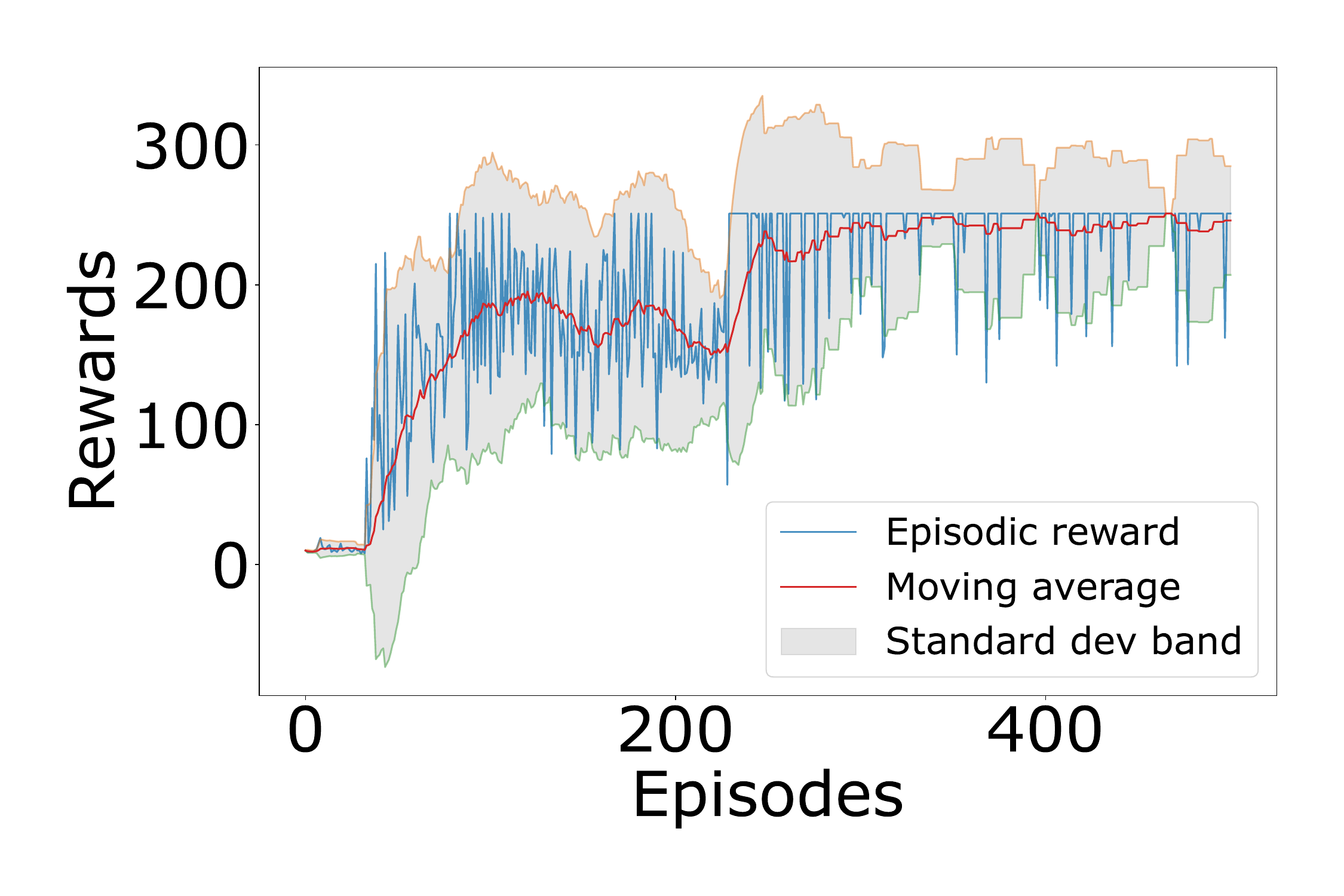}
}
\subfloat[]{
  \includegraphics[clip=true, trim={2cm 2cm 0 1cm},width=5cm]{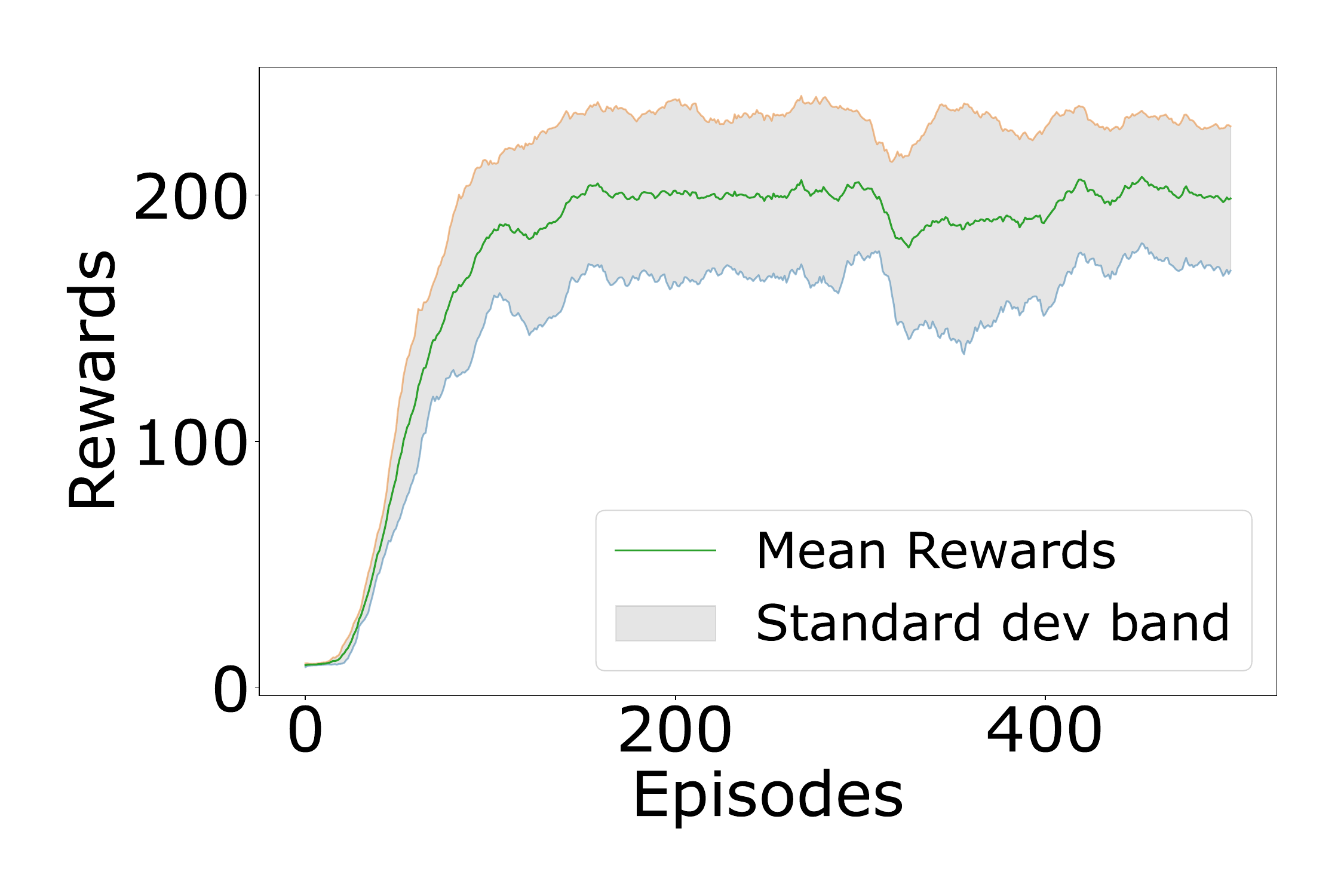}
}
\subfloat[]{
  \includegraphics[clip=true, trim={2cm 2cm 0 0},width=5cm]{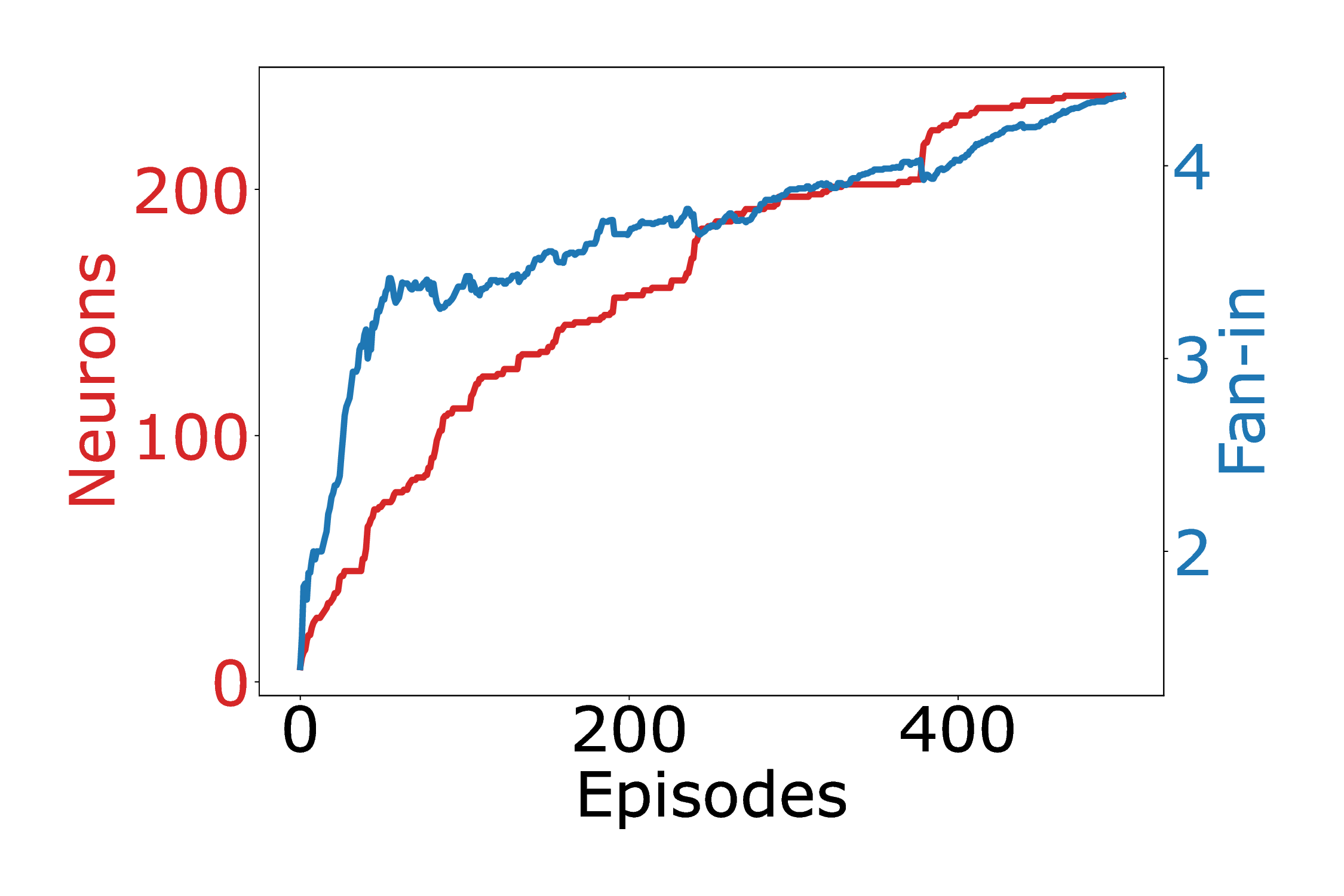}
}
\hspace{0mm}
\subfloat[]{
  \includegraphics[clip=true, trim={2cm 2cm 0 1cm},width=5cm]{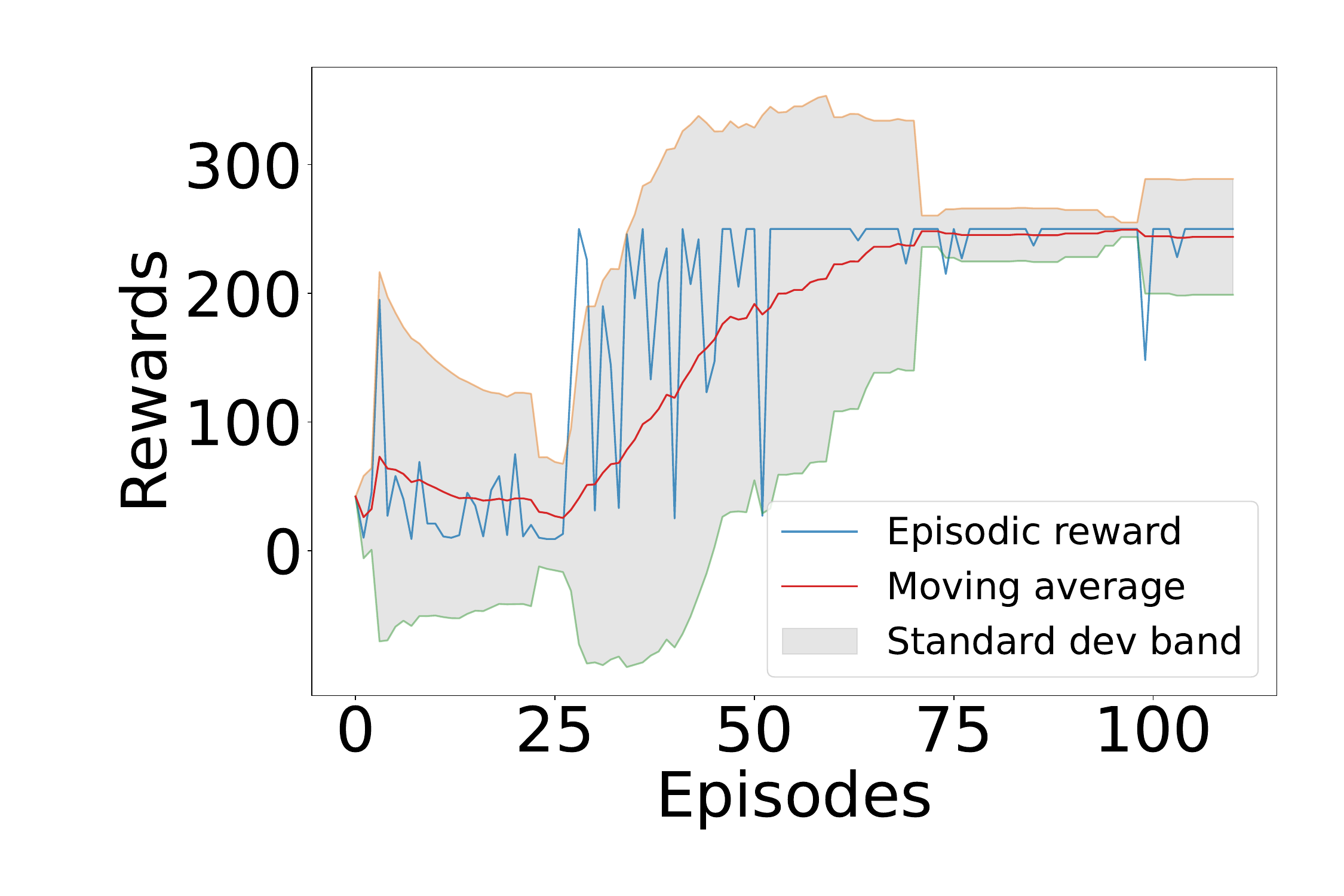}
}
\subfloat[]{
  \includegraphics[clip=true, trim={2cm 2cm 0 1cm},width=5cm]{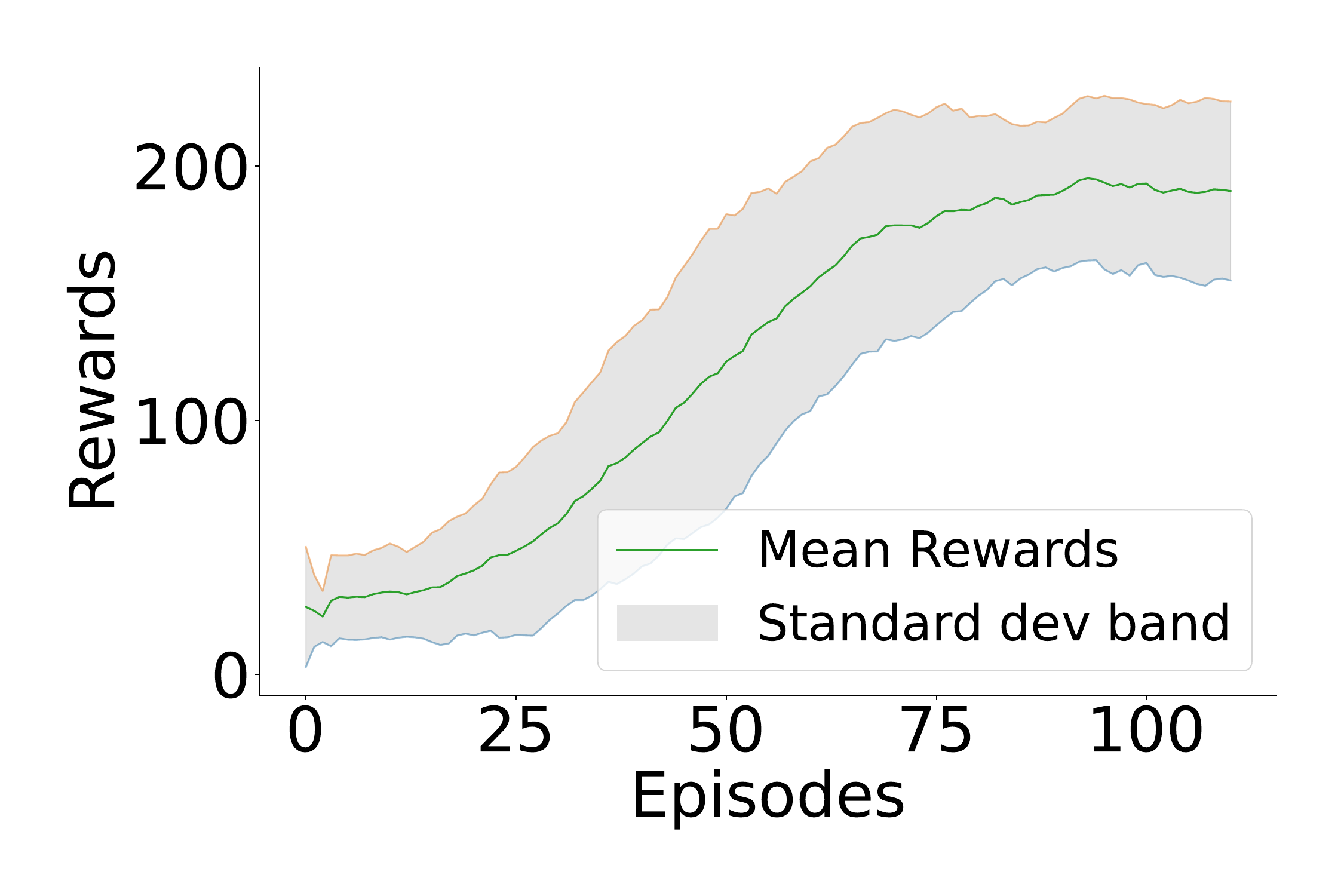}
}
\subfloat[]{
  \includegraphics[clip=true, trim={2cm 2cm 0 0},width=5cm]{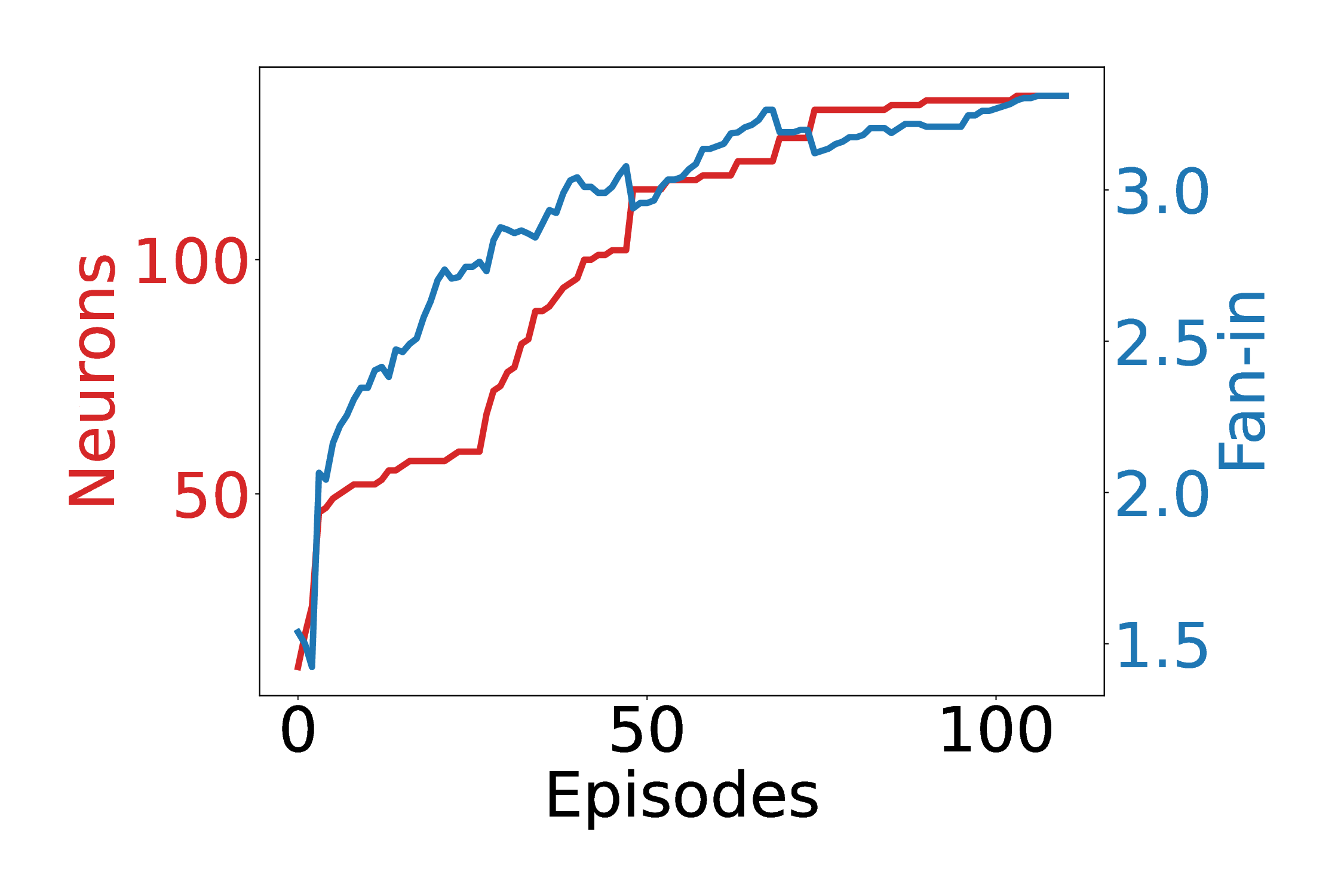}
}
\hspace{0mm}
\subfloat[]{
  \includegraphics[clip=true, trim={2cm 2cm 0 1cm},width=5cm]{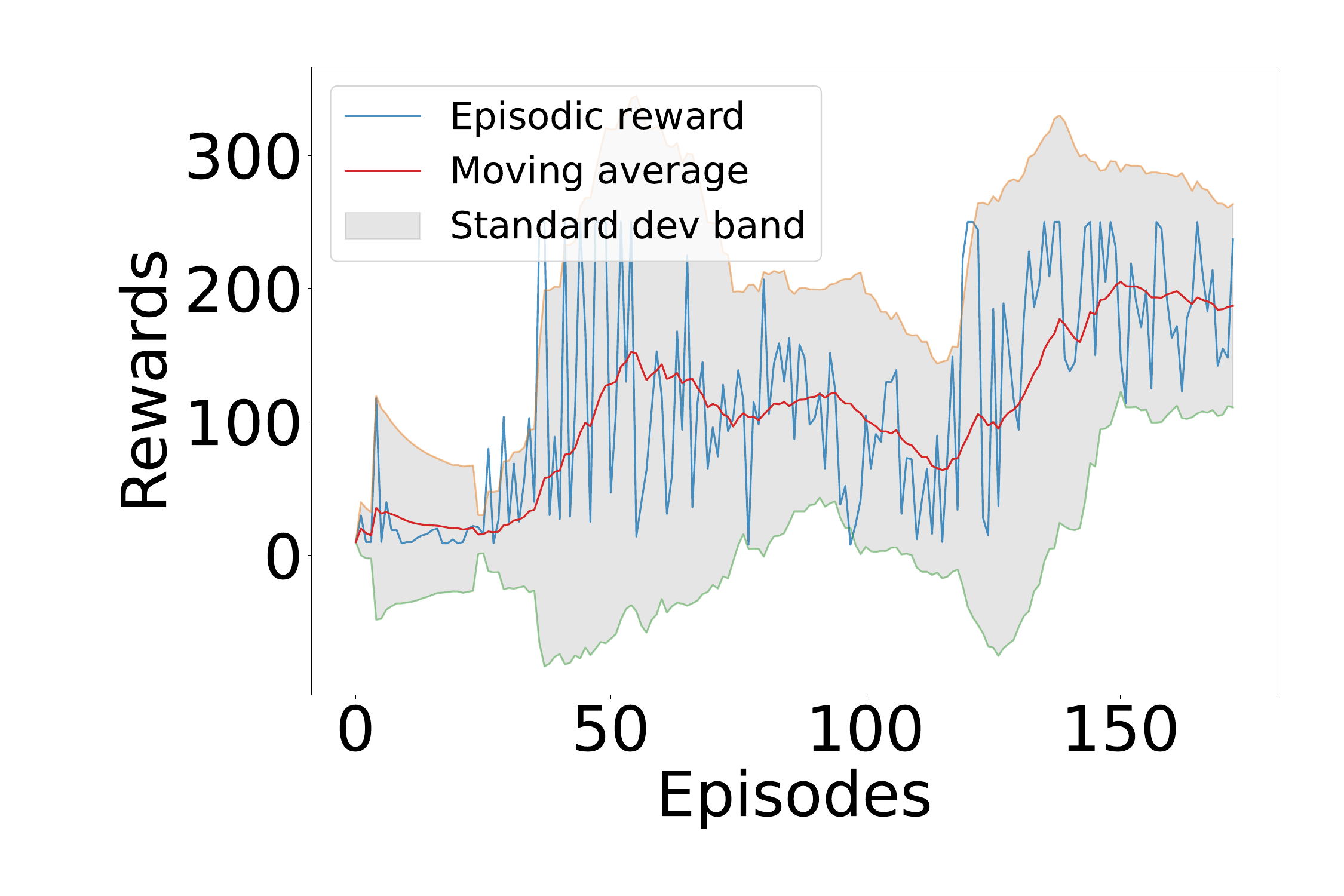}
}
\subfloat[]{
  \includegraphics[clip=true, trim={2cm 2cm 0 1cm},width=5cm]{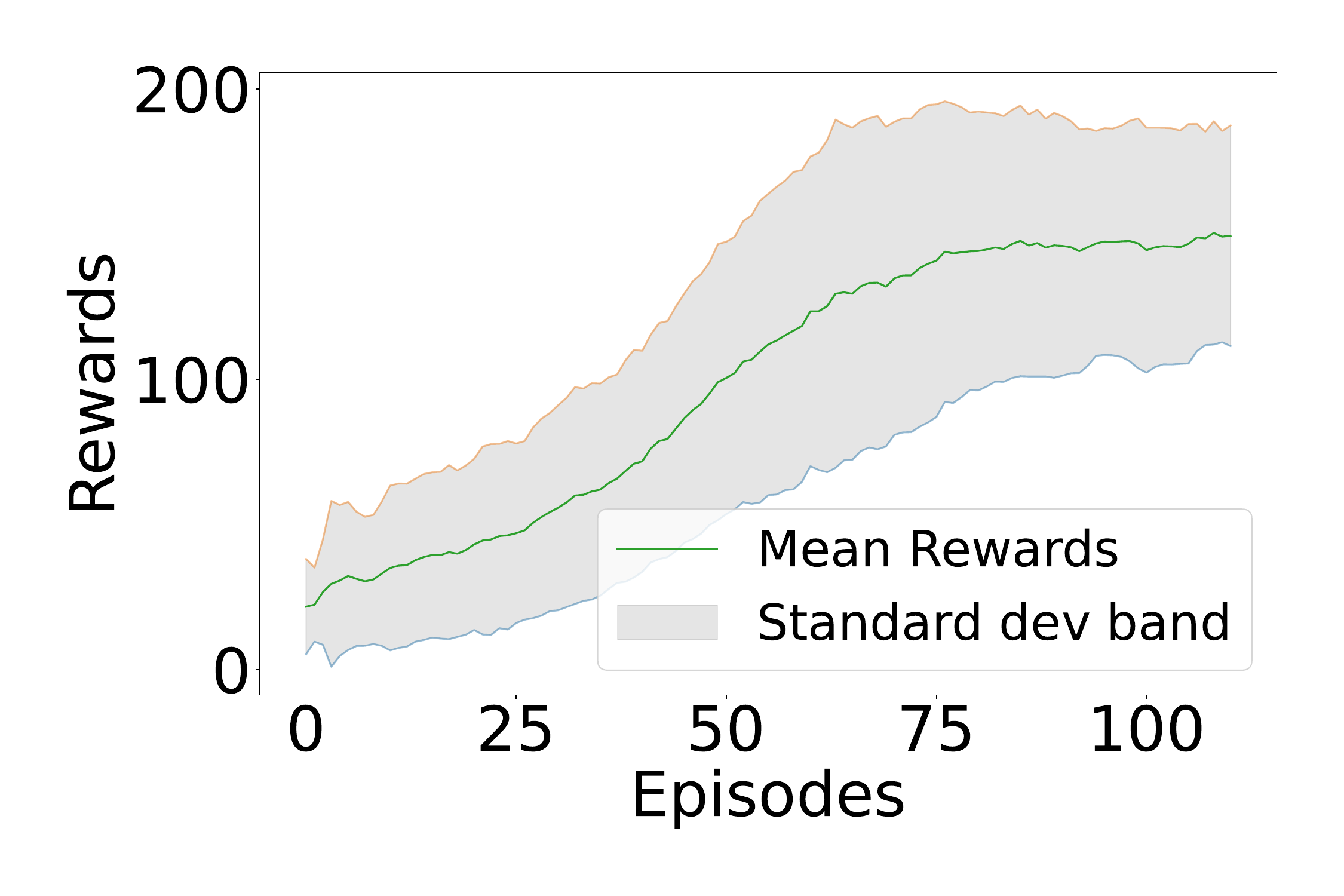}
}
\subfloat[]{
  \includegraphics[clip=true, trim={2cm 2cm 0 0},width=5cm]{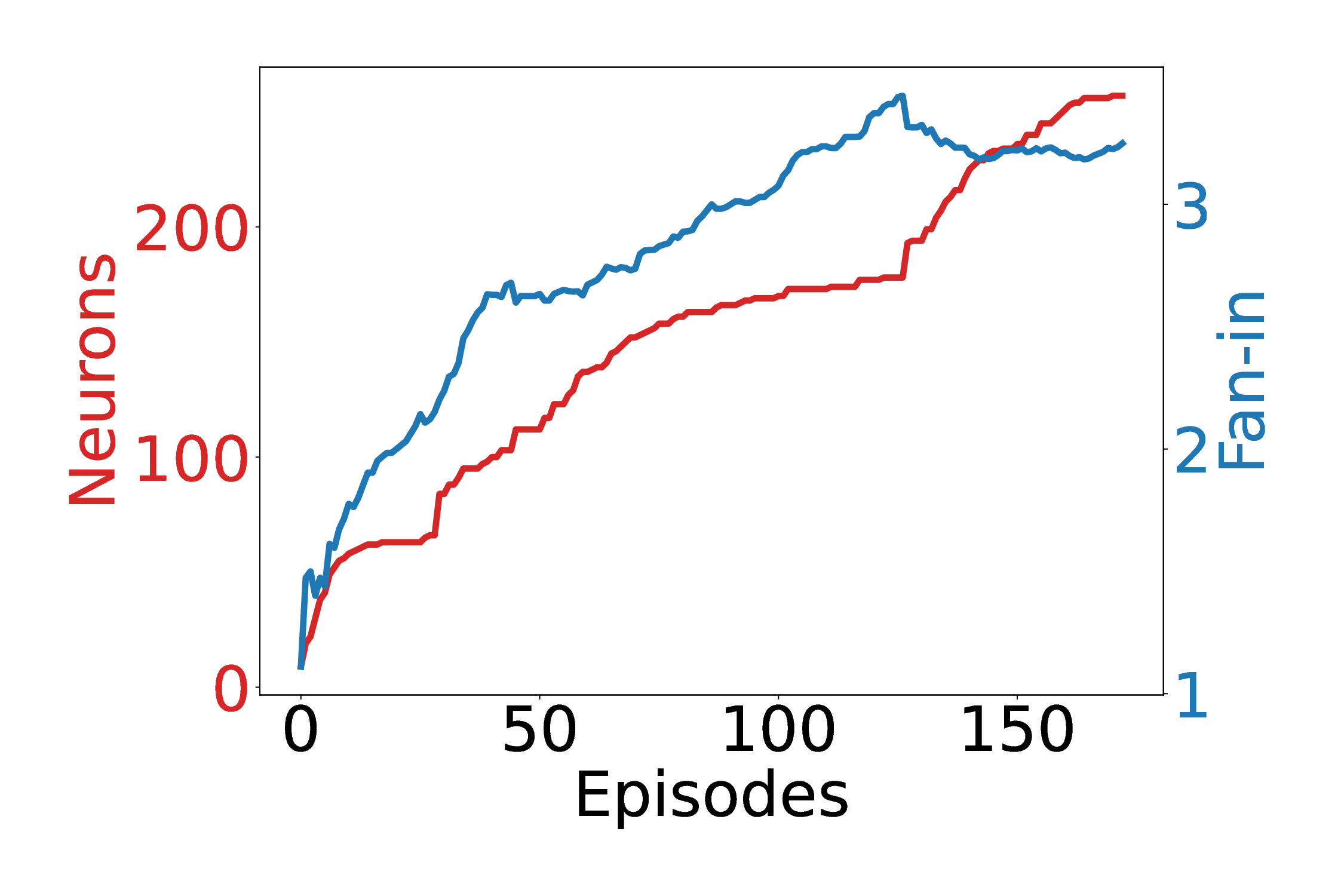}
}
\caption{Plots showing the change of episodic rewards, number of neurons/ensembles, and the fan-in, with training episodes.
Subfigures (a), (d) and (g) show the reward plots for a single training instance for synaptic Q-learning, Bellman memory units using Nengo and using Nengo-Loihi respectively. Subfigures (b), (e) and (h) depict the moving average reward over 10 different training instances for the corresponding cases. Subfigures (c), (f) and (j) illustrate the evolution of network ensembles/neurons with training episodes for the same 3 cases. Each graph is plotted within a band of 2 standard deviations about the mean.}
\label{fig:Bollinger}
\end{figure*}

\section{Training}\label{training}
This algorithm was run and the network trained using a Python IDE, running on a 64 bit Ubuntu 22.04.4 LTS OS in a machine with an Intel\textsuperscript{\textregistered} Core\textsuperscript{\texttrademark} i7-8700K CPU @ 3.70GHz $\times$ 12, with a 32 GB RAM. Training synaptic Q-learning on such a system, took $80$ seconds on an average for $500$ trials.

An example network evolving during training of a synaptic Q-learning model is shown in Fig. \ref{fig:Flow}. As training proceeds, the network topology evolves, but as the training converges, ideally, a single path is followed from every starting state (shown by the open gates), and the rest of the network becomes dormant (closed gates). In Fig. \ref{fig:Flow}, the path through neuron $N_3$ was active at some earlier time instant. But with training, as the value of $Q_{12}(s, a)$ exceeded $Q_{13}(s, a)$, the gate for synapse $S_{13}$ closed and the path was blocked.

For implementation using Nengo, a collection of plots showing the evolution of the network graph over time is shown in Figure \ref{fig:nengo_evolution}. It is seen that the number of ensembles increases and their connectivities become more dense with the training episodes. 
A similar trend is seen for the hardware implementation(see Figure \ref{fig:loihi_evolution}). This implementation is carried out on Intel's Loihi neuromorphic chip, which comprises $256$ cores, each core comprising $1024$ neural compartments.

\begin{table}[htbp]
\caption{Neural network objects for the synaptic Q-learning algorithm and the data variables they store. Objects are along the columns, and the variables are along the rows. 
}
\begin{center}
\begin{tabular}{|c|c|c|c|}
\hline
 & \textbf{Neuron ($\mathcal{N}$)} & \textbf{Synapse ($\mathcal{J}$)} & \textbf{Gate ($\mathcal{G}$)} \\
\hline
\textbf{State} & Value function & Q value & N/A  \\
\hline
\textbf{Connection} & \textcolor{blue}{synapse-synapse}  & \textcolor{blue}{neuron-neuron} & \textcolor{blue}{synapse-neuron} \\
($\overline{c}$)  & & \textcolor{orange}{reward-synapse} &  \\
 &  & \textcolor{orange}{neuron-synapse} &  \\
\hline
\textbf{Fan-in ($\overline{f}$)}   & \checkmark       & N/A      & N/A \\
\hline
\textbf{Kernel} & N/A & $\tau$ & N/A \\
\hline
\end{tabular}
\label{table:symbol}
\end{center}
\end{table}

\section{Results and Discussion}\label{results}

\begin{figure}[htb]
\centering
\subfloat{
  \includegraphics[clip=true, trim={1cm 2cm 0 0},width=7cm]{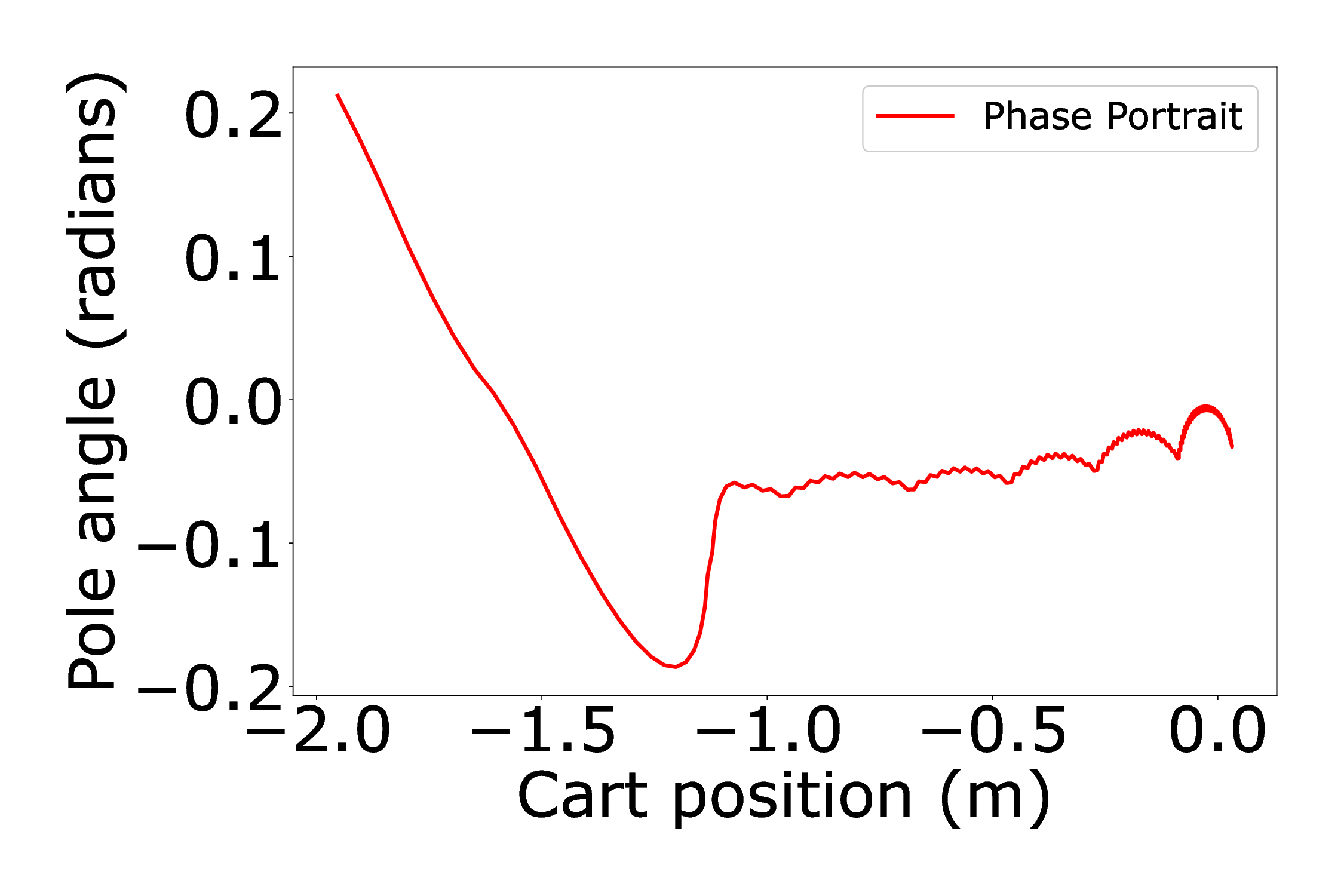}
}
\caption{Phase portrait for trained cartpole using synaptic Bellman equation.}
\vspace{3mm}
\label{fig:Phase}
\end{figure}

\subsection{Synaptic Bellman Equation Implementation}
For $\gamma = 0.99$ and $\alpha = 0.9$,
the reward plot for training synaptic Q-learning algorithm to balance a cartpole has been presented in Fig. \ref{fig:Bollinger}. Fig. \ref{fig:Bollinger}(a) shows the evolution of rewards with training, for a particular training example. Here a window of 20 episodes as per the convergence criterion, is taken to calculate the moving average of the episodic rewards and the standard deviation. It is observed that the controller needs approximately $250$ to $300$ iterations to learn to balance the cartpole. This is acceptable in the context of RL \cite{b8}, as long as the training is repeatable across trials. Fig. \ref{fig:Bollinger}(b) shows the evolution of moving average rewards across $10$ different training instances for robustness checks across trials. In this context, a controller A is considered more robust than a controller B if, across trials, the reward plot for A shows a narrower band of standard deviation about the mean compared to B. The increase in rewards with episodes depicts successful learning for the controller. Fig. \ref{fig:Bollinger}(c) shows the evolution of network topology for a particular training instance. It is seen that initially, the number of neurons and average fan-in per neuron increases as training progresses. However, as more and more regions of the state space get explored with training, this number tends to saturate, implying that the optimal policy $\pi$, and hence the optimal network model is reached. 
Fig. \ref{fig:Phase} shows the phase portrait between the cart position and the pole angle after the training is complete. The system is seen to be controlled within small pole angles about the UEP. After some time, the pole angle crosses $0.2$ radians at a cart position of $-2$ m, and the episode terminates. Since the requirement for RL-based control of the OpenAI cartpole simulator is to maintain the pole within small angles for a certain number of timesteps (ideally, $195$ simulation seconds), the control here is considered to be successful. It is important to note that this control strategy differs significantly from LQR-based control. While LQR asymptotically stabilizes the pole angle toward the unstable equilibrium point (UEP) over time, the phase portrait of the RL-trained cartpole is strongly influenced by the specific reward structure employed.

\begin{figure}[htb]
\centering
\subfloat[]{
  \includegraphics[clip=true, trim={2cm 2cm 0 1cm},width=0.25\textwidth]{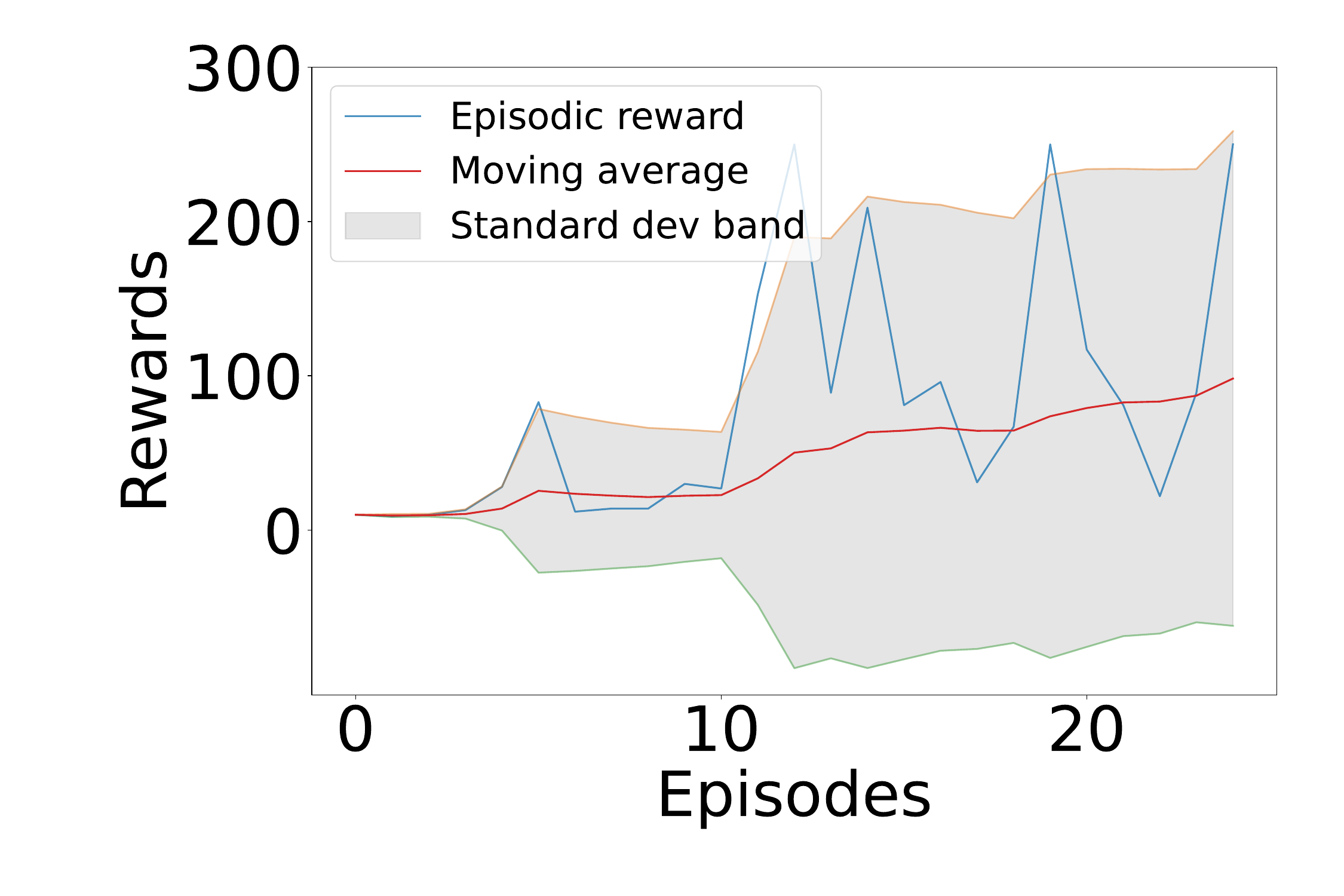}
}
\subfloat[]{
  \includegraphics[clip=true, trim={2cm 2cm 0 0},width=0.25\textwidth]{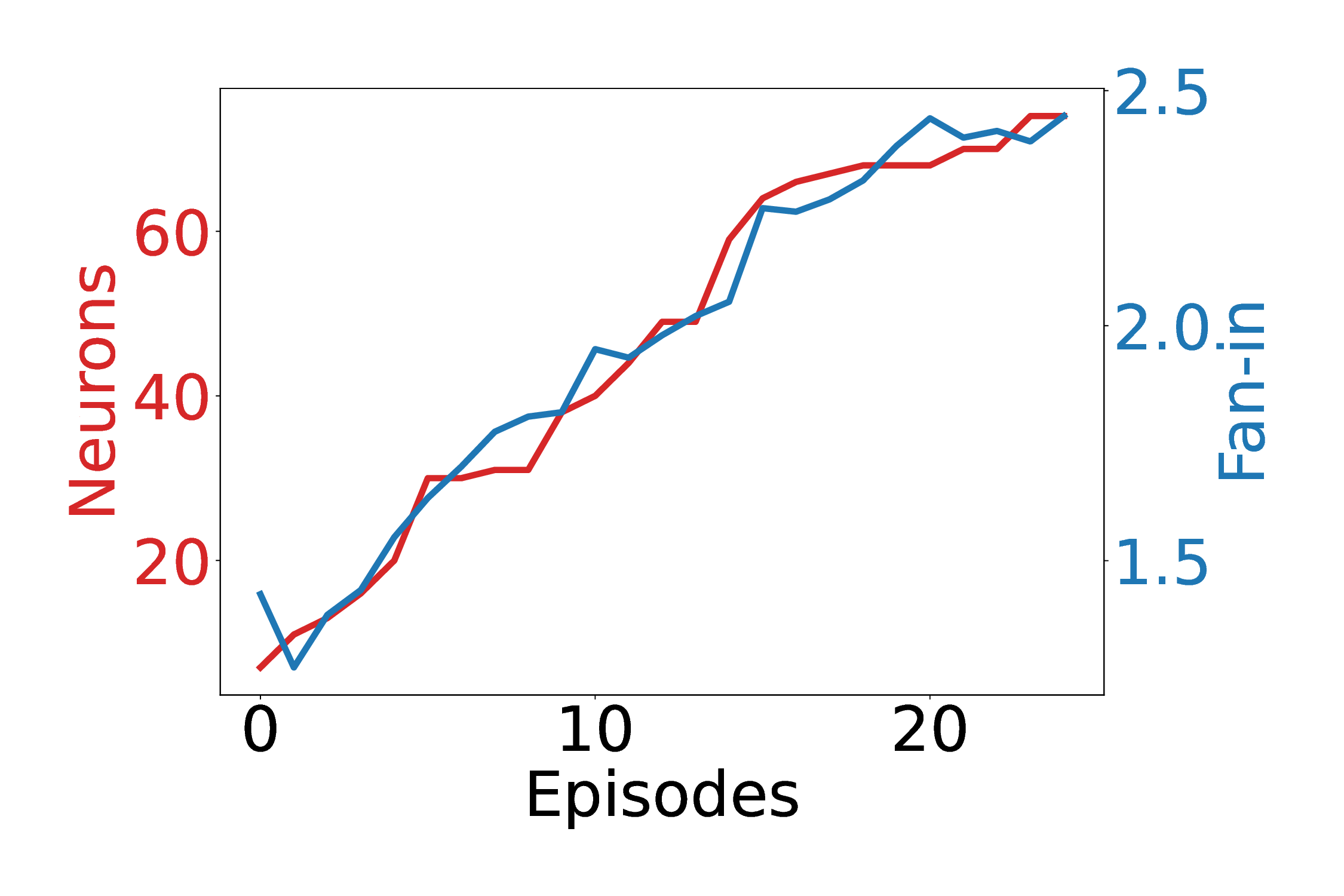}
}
\caption{Training curves for training Bellman memory units on Loihi on a cartpole. (a) shows the reward plots for a single training instance. Each graph is plotted with a band of 2 standard deviations. (b) illustrates the evolution of network topology parameters with training.}
\label{fig:Bollinger_loihi}
\end{figure}

\begin{figure}[htb]
\centering
\subfloat[]{
  \includegraphics[clip=true, trim={2cm 2cm 0 1cm},width=0.25\textwidth]{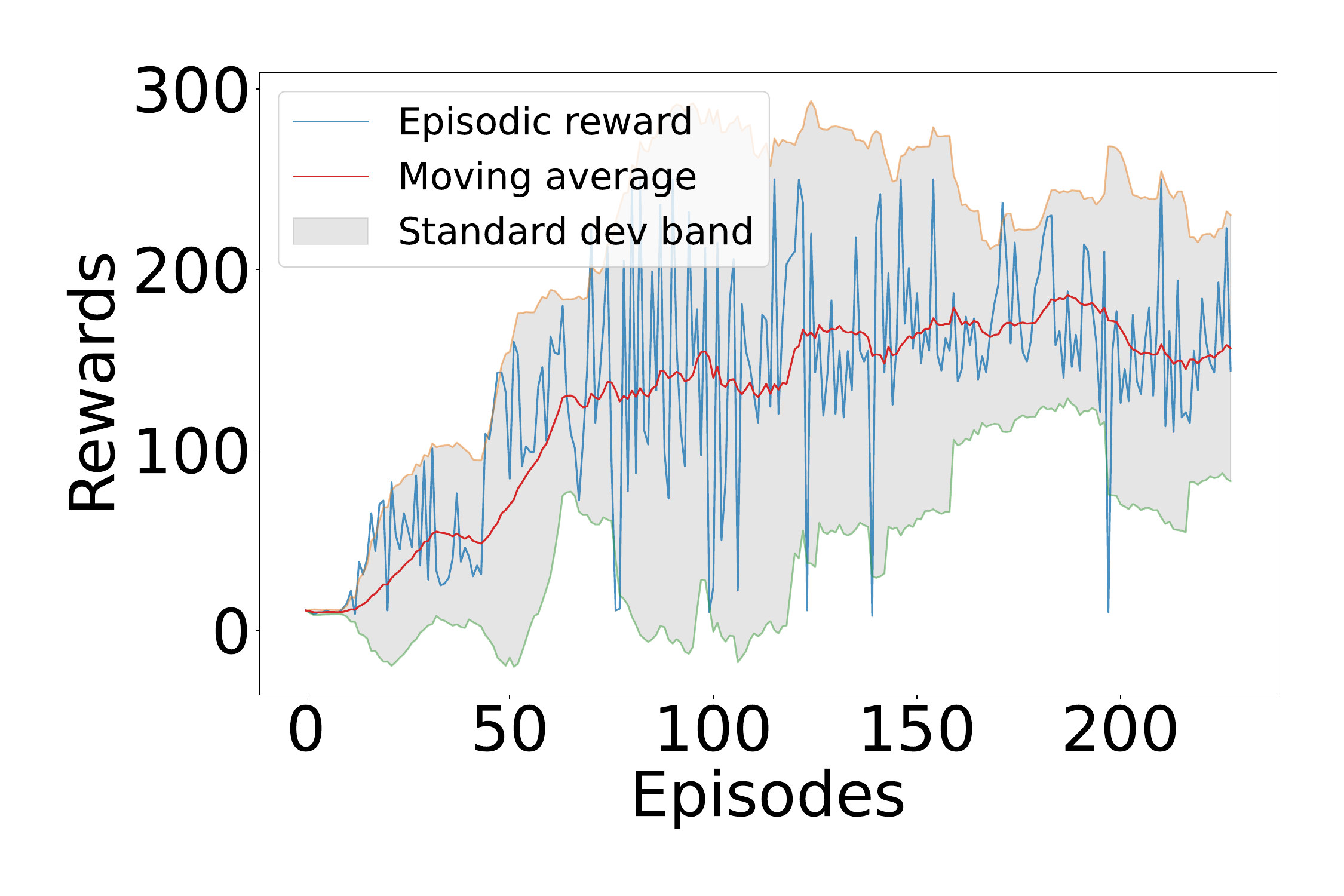}
}
\subfloat[]{
  \includegraphics[clip=true, trim={2cm 2cm 0 0},width=0.25\textwidth]{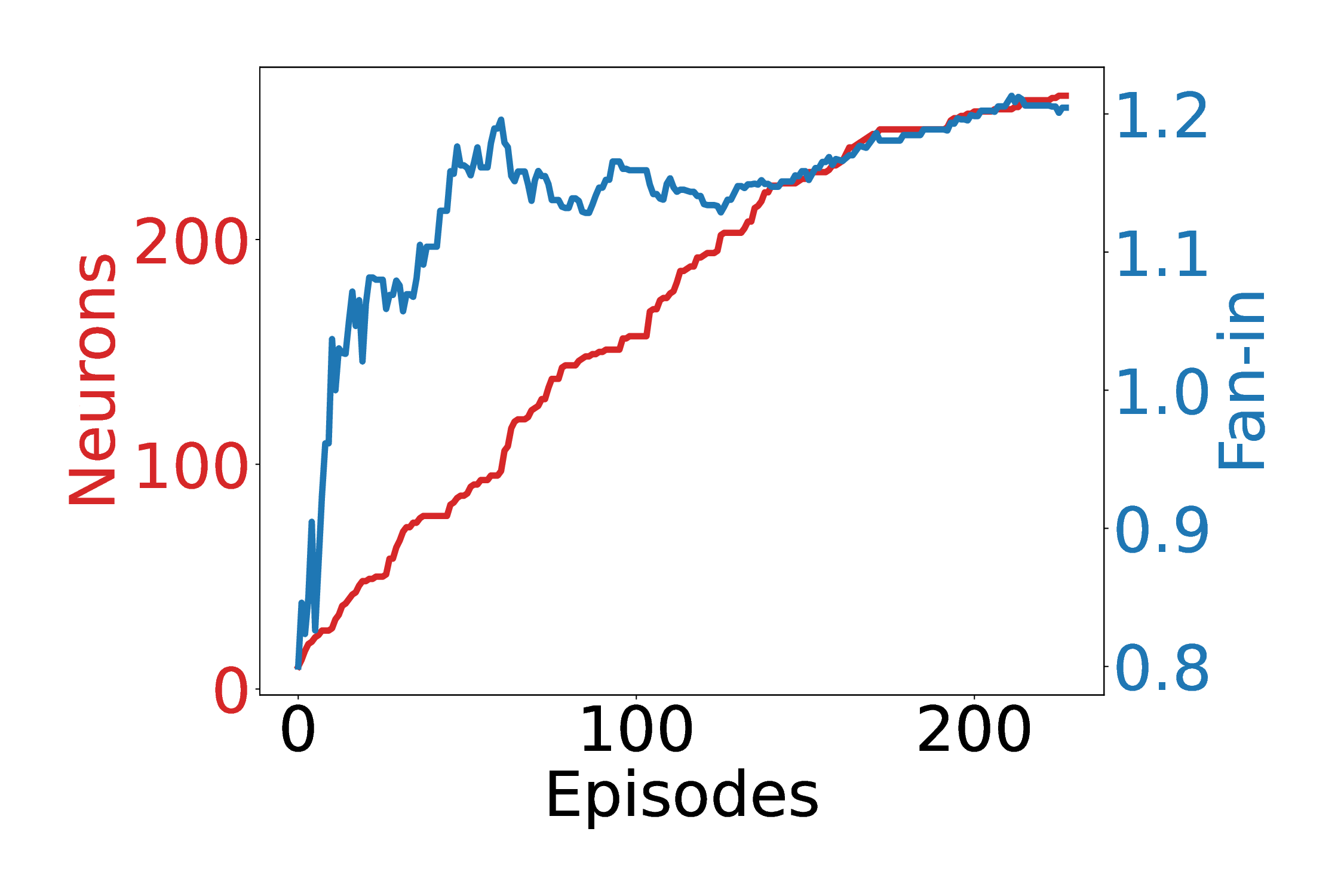}
}
\caption{Training curves for training Bellman memory units using a single ensemble on Loihi on a cartpole. (a) shows the reward plots for a single training instance. Each graph is plotted with a band of 2 standard deviations. (b) illustrates the evolution of network topology parameters with training.}
\vspace{3mm}
\label{fig:Bollinger_loihi_1}
\end{figure}

\begin{table*}[htb]
\small
\caption{Comparison between synaptic Q-learning, Bellman memory units, DDQN and Q-table methods of learning.}
\begin{center}
\begin{tabular}{|c|c|c|c|c|c|c|}
\hline
 & \textbf{Synaptic}& \textbf{Bellman}& \textbf{Bellman}& \textbf{Bellman} & \textbf{Q-Table} & \textbf{DDQN} \\
  & \textbf{Q} & \textbf{memory} & \textbf{memory}& \textbf{memory} &  &  \\
   & \textbf{(Python)} & \textbf{(Nengo-CPU)} & \textbf{(Nengo-Loihi)} & \textbf{(Loihi)}&  &  \\
\hline
\textbf{Episodes} & $250$ & $100$ & $200$ & $200$ &$310$ & $250$ \\
\hline
\textbf{Neurons} & $250$ & $140$ &  $250 $& $300$ & $1e4$ & $600$\\
\hline
\textbf{Fan-in}   & $4$ &  $3.25$ &  $3.5$ & $1.2$ & $4$  & $56$ \\
\hline
\textbf{Parameters}  &   $1000$  & $455$ & $875$ & $360$ & $40000$ & $33600$\\
\hline
\end{tabular}
\label{table:metrics}
\end{center}
\end{table*}

\subsection{Nengo and Nengo-Loihi Implementation}
The synaptic Q-learning algorithm was also implemented using the NEF (Algorithm \ref{Algo:2}) to investigate the applicability and benefits of using such a framework for this algorithm.

\subsubsection{Network graph results}

\begin{figure}[ht!]
\centering
\includegraphics[width = 0.4\textwidth, trim = {0cm 5cm 0cm 5cm}, clip]{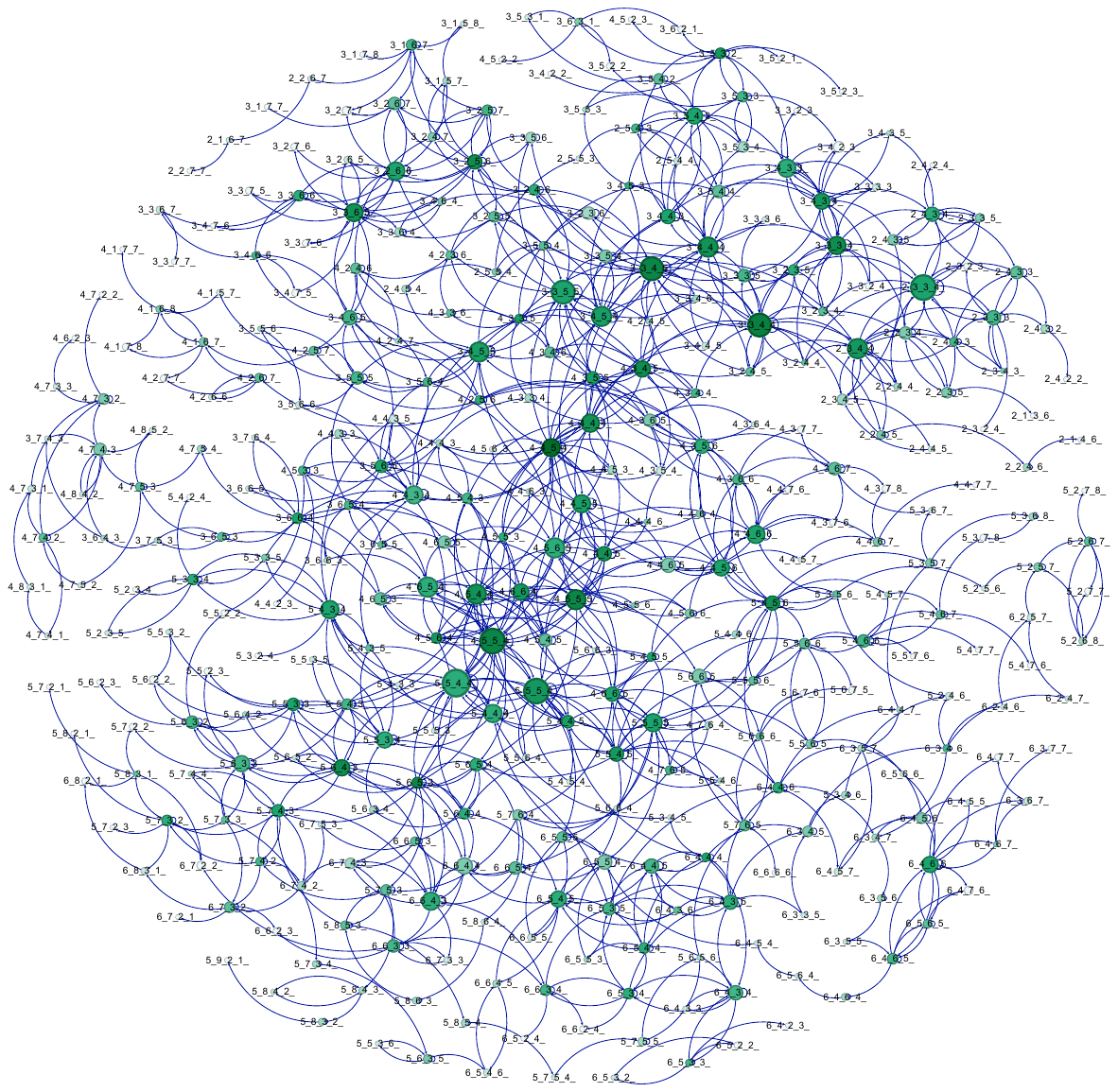}
\caption{Final network graph for training the synaptic Bellman equation using Nengo on CPU. The circles denote the nodes and ensembles, and the lines show the interconnections among them. The size of a circle is proportional to the number of incoming connections, and the opacity of the circle is proportional to the number of outgoing connections.}
\vspace{3mm}
\label{fig:Gephy_network_topology}
\end{figure}

Figure \ref{fig:Gephy_network_topology} shows the final network graph upon training Bellman memory units with Nengo on CPU. Each blob denotes an ensemble or node spawned based on the discretized value of a cartpole state, visible on zooming into the plot. The size of a blob is proportional to the number of incoming connections, and the opacity of the blob denotes the number of outgoing connections from the ensemble. The lines with arrow-heads denote the interconnections between the ensembles. 

Here, the value of every state variable ($x$, $\dot{x}$, $\theta$, and $\dot{\theta}$) spans between some negative and positive value, and is symmetrical about zero (See Figure \ref{fig:DisFlow}). Considering $10$ discretization bins, $0$ is assigned to the lowest negative value, and $9$ is assigned to the highest positive value of the state variable. Thus, $[5,5,5,5]$ is considered to be the discretized version of the state $[0.0,0.0,0.0,0.0]$. 

It is important to note that the larger and more opaque blobs have their $\theta$ and $\dot{\theta}$ values close to $0$ radians, while the smaller and transparent blobs have their corresponding values further away from $0$. This indicates that, for the trained policy, the states visited more are those close to the UEP, and hence the training brings the pole angle closer to the UEP over time.

Due to the chosen precision of discretization, closely-spaced consequent states are treated as the same state in their discretized form, connecting a blob to itself. That is why multiple blobs show recurrent connections in the graph (Figures \ref{fig:Gephy_network_topology} and \ref{fig:nengo_evolution}).

Moreover, there are visible clusters in the graph arising due to the interconnectivities. States that appear more often in the final controlled trajectories have their blobs more densely interconnected compared to the others and are more closely spaced. This results in the formation of clusters. 

Some ensemble pairs are interconnected as a loop, demonstrating an oscillatory behaviour between those states. For example, ensemble 3\textunderscore4\textunderscore6\textunderscore5\textunderscore  has a connection loop with ensemble 3\textunderscore3\textunderscore6\textunderscore5\textunderscore, which in turn has one with ensemble 3\textunderscore2\textunderscore6\textunderscore5\textunderscore. Here, the value of pole angle is positive (close to $\approx 0.0836$ radians), while the angular velocity is close to $0$. Hence, a left acceleration on the cart is necessary to bring the pole angle back to $0$. However, the acceleration shouldn't be continuously high to prevent the pole from crossing the UEP and moving to negative angles. As a result, an oscillatory left-right push is applied to the cart, causing the cart's velocity to oscillate while making incremental changes to the pole angle and angular velocity within the same discretization bin.

Additionally, the biggest and most opaque blobs have their angles and angular velocities close to $0$ radians. This clearly indicates that after the training is complete, the control trajectories hover about the pole's UEP.

\begin{figure}[ht!]
\centering
\includegraphics[width = 0.3\textwidth]{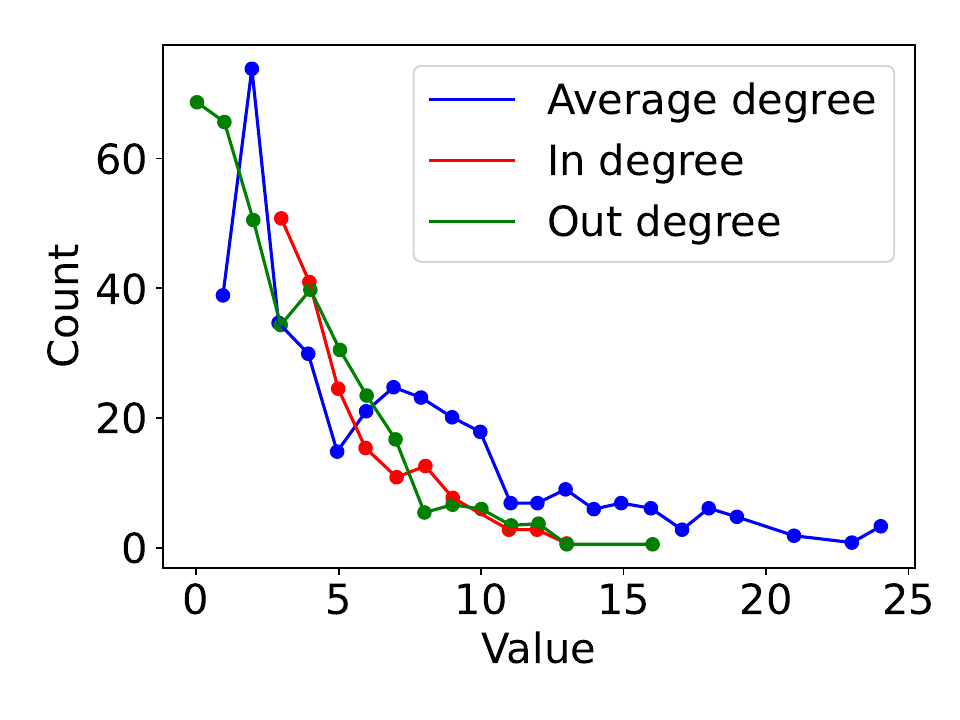}
\caption{Degree distribution plot for a trained graph when Bellman memory units are trained using Nengo on CPU.}
\label{fig:Degree_distribution}
\end{figure}

Figure \ref{fig:Degree_distribution} shows the degree distribution of the trained graph when Bellman memory units are trained on a CPU with Nengo. The degree distribution represents the number of connections per node (x-axis) versus the number of nodes with that degree (y-axis). It shows that, though a maximum of 80 ensembles have a degree of about $3$, and the average degree is about $3.157$, the degree can reach a maximum of $25$. As a result, a minimum provision for $25$ connections must be allowed per ensemble, when designing a neuromorphic ASIC for this application. The relatively low average degree further indicates sparse connectivity, a fundamental characteristic of neuromorphic architectures that contributes to their low power consumption.

\subsubsection{Training results}
The training curves showing the evolution of rewards over time, the number of ensembles, and the fan-in are illustrated in Figure \ref{fig:Bollinger}(d),(e) and (f).


 Clearly, the training time for Bellman memory units is around $100$ epochs while that for synaptic Bellman equation is $250$ epochs. However, a robustness check shows the Nengo implementation saturating at a lower value of maximum reward as compared to the Python implementation, which implies that training for Bellman memory units is faster at the cost of lower robustness. The number of neurons and the average fan-in are comparable for both cases.

Figure \ref{fig:Bollinger}(g),(h),and(i) show the corresponding plots for online training of Bellman memory units on Nengo-Loihi. This implementation is much poorer than the Nengo simulation, both in terms of robustness and convergence. However, the convergence of Nengo-Loihi Bellman memory units is still better than the synaptic Bellman equation in Python. Figure \ref{fig:Bollinger_loihi} shows the Bollinger reward plot and the network parameter evolution plot for training on the Loihi neuromorphic chip. The results are incomplete as the simulation aborted after running for about $25$ iterations. This is because, in the Nengo implementation, a distinct ensemble is assigned to each new state encountered. When running the same algorithm on Loihi with Nengo-Loihi, a separate Loihi chip is allocated for each ensemble. Hence, even though there are only 2 neurons per ensemble, and the per-block/chip utilization is only $1.4\%$, overall, the number of chips required by Loihi exceeds the maximum available on the board, which is $256$ in our case. This can be seen from the Loihi resource utilization summary as shown below:

\begin{figure}[ht!]
\centering
\includegraphics[width = 0.5\textwidth]{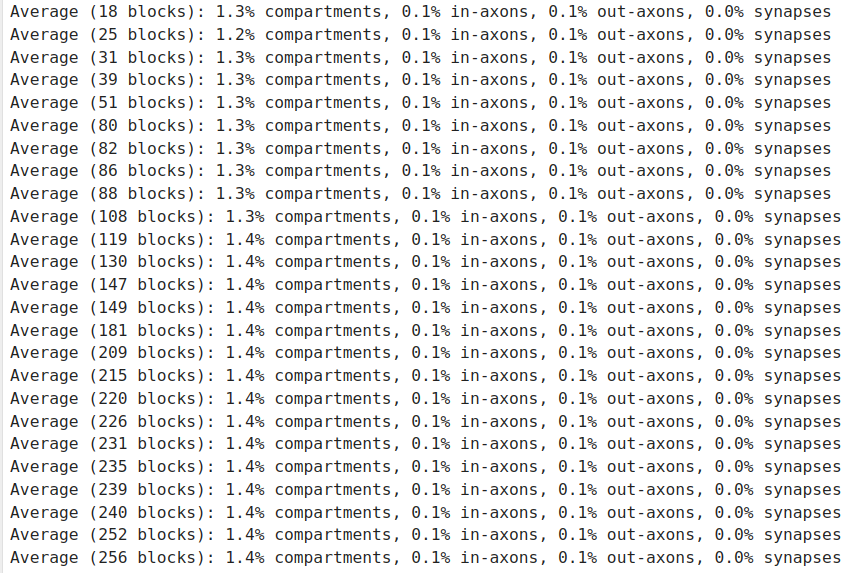}
\caption*{Dynamic allocation of Loihi chips and compartments during learning to balance a cartpole using Nengo-Loihi.}
\end{figure}


\subsection{Loihi Implementation}
To implement this algorithm on Loihi, without causing the board to run out of resources, a single ensemble-based implementation is adopted. On the Loihi board implementation, a particular neuron in the ensemble represents a distinct state, as opposed to assigning a full ensemble to a state in Nengo-Loihi. Based on some prior knowledge obtained from the Nengo-Loihi implementation, it is seen that around $250$ new ensembles are spawned. Using some factor of safety, a $300$ neuron ensemble is initialized for this control task. Nengo does not allow for the dynamic addition of neurons to an existing ensemble. So the only option is to start with an ensemble containing a certain number of neurons matching an upper-estimate, based on prior knowledge. As the simulation progresses, a distinct neuron in the ensemble is assigned to a new state. If the action contains $n$ discretization bins, $n$ connectivities (axons) are spawned from the neuron. Based on the chosen action, the corresponding axon is connected to the neuron assigned to the next state via a Nengo node. The transformation weight on this axon is updated with the Q-value corresponding to the state-action pair. Whenever a state is encountered, the chosen neuron sends spikes through its $n$ axons. The axon with the maximum weight, sends the maximum pre-synaptic current, and that specific action is selected. Based on the trials conducted, an ensemble initialized with $600$ neurons was found to be adequate. This is much less compared to the $1024$ neural compartments provided per chip in Loihi. The corresponding resource utilization on Loihi is $58.6\%$ of a chip as seen from the utilization summary.

\begin{figure}[ht!]
\centering
\includegraphics[width = 0.5\textwidth]{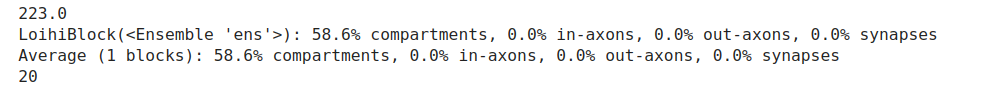}
\caption*{Resource utilization on Loihi for 1 ensemble-based control of a cartpole using Bellman memory units.}
\label{fig:utilization}
\end{figure}

Figure \ref{fig:Bollinger_loihi_1} illustrates the reward plot and the network topology evolution plot for a single ensemble-based control of a cartpole using Bellman memory units implemented on Loihi. The reward plot shows a large standard deviation during training, which implies that training on Loihi is not robust. Clearly, the number of neurons spawned is more compared to the Nengo and Nengo-Loihi implementations, but the fan-in per neuron is approximately $33\%$ of the aforementioned implementations.

\subsection{Comparison with other methods}
Table \ref{table:metrics} shows a comparison of synaptic Q-learning and Bellman memory units, with a standard Q-table-based approach, and a DDQN-based learning for stabilizing a cartpole. For the Q-table, a bin size of 10 is taken for each component of the state, similar to the proposed algorithm. A key advantage of the proposed approach is the need for a reduced number of neurons and connections, described by the fan-in, which subsequently reduces the number of memory elements and hence, the size on chip. Training time for synaptic Q-learning is comparable to the other methods. Also, this approach operates on spikes, and hence, it has the potential to be implemented on a neuromorphic chip. The implementation on Nengo shows quicker convergence and much less network parameters than the other methods. However, Nengo-Loihi and Loihi implementations show poorer performance compared to Nengo simulation. One advantange of single ensemble implementation on Loihi is the reduced fan-in, and hence the lower number of network parameters compared to other methods.

\section{Conclusion}\label{conclusion}
This work presents (a) a synaptic Q-learning algorithm to balance a cartpole in its UEP using Python, (b) Bellman Memory Units on CPU using Nengo, (c) Bellman Memory Units on CPU with a Loihi-like simulator using Nengo-Loihi, and (d) the implementation of single ensemble-based Bellman Memory Units on Loihi.
The synaptic Q-learning algorithm network iteratively evolves as learning progresses. Due to the mixed analog-digital form of computation, and derivative-free learning, this architecture demonstrates the potential to be used with neuromorphic chips.
Results indicate that learning converges in $200$ to $250$ training episodes, and is robust across trials. Also, the network topology stabilizes as the learning converges, yielding an optimal number of neurons and synapses required for a particular control application. This becomes useful for optimizing on-board resource utilization, which may subsequently reduce chip size and power consumption. This algorithm is followed by a similar approach called the BMU implemented using the Nengo NEF simulator, on the Nengo-Loihi simulator, and on Intel's Loihi neuromorphic chip. Although the on-chip implementation is less robust than the Nengo simulator, it still performs better than some conventional RL algorithms like Q-table and DDQN, in terms of training time and network parameters needed. BMUs, when implemented on Loihi, require $2$ orders of magnitude less memory elements than a conventional Q-table, and half the number of neurons required by a standard DDQN. The Loihi implementation with a single ensemble requires $33\%$ of synaptic connections when compared with implementation of Algorithm \ref{Algo:1}, Nengo, and Nengo-Loihi implementations.  The main advantage of this method is the on-chip implementation and online learning of a memory-based neuromorphic decision-making algorithm, allowing for adaptability in control.

This work only considers the greedy version of the synaptic Q-learning algorithm presented, which does not include randomness in action selection to explore the environment. This can lead to problems like the training getting stuck in undesirable local optima. To enable exploration-exploitation trade-off, the gate may be designed to have a probabilistic operation rate where the probability of random action selection decreases as the training progresses.

\section*{Acknowledgment}
This publication has emanated from research conducted with the financial support of Taighde Éireann – Research Ireland under Grant number 18/CRT/6049. For the purpose of Open Access, the author has applied a CC BY public copyright license to any Author Accepted Manuscript version arising from this submission. Vikram Pakrashi acknowledges Research Ireland NexSYs 21/SPP/3756, Research Ireland I-Form Advanced Manufacturing Research Centre,
and Sustainable Energy Authority of Ireland REMOTEWIND
RDD/613, TwinFarm RDD/604, FlowDyn RDD/966 and Interreg SiSDATA EAPA-0040/2022.


\begin{thebibliography}{99}
\bibitem{b1} C. D. Schuman, S. R. Kulkarni, M. Parsa, J. P. Mitchell, and B. Kay. "Opportunities for neuromorphic computing algorithms and applications." Nature Computational Science 2, no. 1 (2022): 10-19.
\bibitem{b2} T. DeWolf, P. Jaworski, and C. Eliasmith. Nengo and low-power ai hardware for robust, embedded neurorobotics. Frontiers in Neurorobotics, 14:568359, 2020.
\bibitem{b3} G. Gallego, T. Delbrück, G. Orchard, C. Bartolozzi, B.Taba, A.  Censi, S. Leutenegger et al. "Event-based vision: A survey." IEEE transactions on pattern analysis and machine intelligence 44, no. 1 (2020): 154-180.
\bibitem{b4} Y. Yang, et al. "Neuromorphic electronics for robotic perception, navigation and control: A survey." Engineering Applications of Artificial Intelligence 126 (2023): 106838.
\bibitem{b5} C. Bartolozzi, G. Indiveri, and E. Donati. "Embodied neuromorphic intelligence." Nature communications 13, no. 1 (2022): 1024.
\bibitem{b6} A. Bhargava, M. R. Rezaei, and M. Lankarany. "Gradient-Free Neural Network Training via Synaptic-Level Reinforcement Learning." AppliedMath 2, no. 2 (2022): 185-195.
\bibitem{b7} H. Qiu, et al. "Evolving spiking neural networks for nonlinear control problems." 2018 IEEE Symposium Series on Computational Intelligence (SSCI). IEEE, 2018.
\bibitem{b8} A. S. Lele, et al. "Learning to walk: Bio-mimetic hexapod locomotion via reinforcement-based spiking central pattern generation." IEEE Journal on Emerging and Selected Topics in Circuits and Systems 10.4 (2020): 536-545.
\bibitem{b9} A. Lahgere. "Design of leaky integrate and fire neuron for spiking neural networks using trench bipolar I-MOS." IEEE Transactions on Nanotechnology (2023).
\bibitem{b10} Mahmoud A., et al. "Porting deep spiking q-networks to neuromorphic chip loihi." International Conference on Neuromorphic Systems 2021. 2021.
\bibitem{b11} R. S. Sutton, and A. G. Barto. "Reinforcement learning: An introduction." MIT press, 2018.
\bibitem{b12} Y. Mothanna, and N. Hewahi. "Review on reinforcement learning in cartpole game." 2022 International Conference on Innovation and Intelligence for Informatics, Computing, and Technologies (3ICT). IEEE, 2022.
\bibitem{b13} E.Raponi, N. C. Rakotonirina, J. Rapin, C. Doerr, and O. Teytaud. "Optimizing with low budgets: A comparison on the black-box optimization benchmarking suite and openai gym." IEEE Transactions on Evolutionary Computation (2023).
\bibitem{b14} Trevor Bekolay, James Bergstra, Eric Hunsberger, Travis DeWolf, Terrence Stewart, Daniel
Rasmussen, Xuan Choo, Aaron Voelker, and Chris Eliasmith. Nengo: a python tool for building
large-scale functional brain models. Frontiers in Neuroinformatics, 7, 201.
\bibitem{b15} Davies, M., Srinivasa, N., Lin, T.H., Chinya, G., Cao, Y., Choday, S.H., Dimou, G., Joshi, P., Imam, N., Jain, S. and Liao, Y., 2018. Loihi: A neuromorphic manycore processor with on-chip learning. Ieee Micro, 38(1), pp.82-99.
\bibitem{b16} Shrestha, A., Fang, H., Mei, Z., Rider, D.P., Wu, Q. and Qiu, Q., 2022. A survey on neuromorphic computing: Models and hardware. IEEE Circuits and Systems Magazine, 22(2), pp.6-35.
\bibitem{b17} Safa, A., 2024. Continual learning with Hebbian plasticity in sparse and predictive coding networks: a survey and perspective. Neuromorphic Computing and Engineering, 4(4), p.042001.
\bibitem{b18} Kaplanis, C., Shanahan, M. \& amp; Clopath, C.. (2018). Continual Reinforcement Learning with Complex Synapses. Proceedings of the 35th International Conference on Machine Learning, in Proceedings of Machine Learning Research. 80:2497-2506. Available from https://proceedings.mlr.press/v80/kaplanis18a.html.



\end{thebibliography}
\end{document}